\documentclass[a4paper,twocolumn, iop, twocolappendix, numberedappendix]{oja}
\usepackage{graphicx} 
\usepackage[T1]{fontenc}
\usepackage[utf8]{inputenc}
\usepackage{natbib}
\usepackage{fontawesome5}
\usepackage{multirow}
\usepackage{comment}
\usepackage{graphicx}	
\usepackage{amssymb}	
\usepackage{xspace}        
\usepackage[dvipsnames]{xcolor}
\usepackage{newtxtext,newtxmath}
\usepackage{booktabs}
\usepackage{fancyhdr}
\usepackage{upgreek}
\usepackage{tensor}
\usepackage{times}
\usepackage{lastpage}
\usepackage{caption}
\usepackage{subcaption}
\usepackage{tikz}
\usetikzlibrary{angles, quotes, 3d}
\usepackage{newtxtext,newtxmath}

\usepackage[colorlinks,linkcolor=blue,citecolor=blue,urlcolor=blue ]{hyperref}
\usepackage{xurl} 

\newcommand{\review}[1]{\textcolor{black}{#1}}

\usepackage{cleveref}

\begin{document}
\title{Baryonification III: an accurate analytical model for the dispersion measure probability density function of Fast Radio Bursts}
\author{MohammadReza Torkamani$^{\dagger}$, Robert Reischke$^{\star}$}
\affiliation{Argelander-Institut für Astronomie, Universität Bonn, Auf dem Hügel 71, D-53121 Bonn, Germany}
\thanks{$^\dagger$\href{mailto:mohamadrezatorkamani@gmail.com}{mohamadrezatorkamani@gmail.com}\\
$^\star$\href{mailto:reischke@posteo.net}{reischke@posteo.net}, \href{mailto:rreischke@astro.uni-bonn.de}{rreischke@astro.uni-bonn.de}}

\author{Michael Kova\v{c}}
\affiliation{Jodrell Bank Centre for Astrophysics, Department of Physics and Astronomy, The University of Manchester, Manchester M13 9PL, United Kingdom \\
Argelander-Institut für Astronomie, Universität Bonn, Auf dem Hügel 71, D-53121 Bonn, Germany}

\author{Andrina Nicola}
\affiliation{Jodrell Bank Centre for Astrophysics, Department of Physics and Astronomy, The University of Manchester, Manchester M13 9PL, United Kingdom}

\author{Jozef Bucko, Alexandre Refregier}
\affiliation{Institute for Particle Physics and Astrophysics, ETH Zurich, Wolfgang Pauli Strasse 27, 8093 Zurich, Switzerland}

\author{Sambit K. Giri}
\affiliation{Department of Astronomy and Oskar Klein Centre, AlbaNova, Stockholm University, SE-10691 Stockholm, Sweden}

\author{Aurel Schneider}
\affiliation{Department of Astrophysics, University of Zurich, Winterthurerstrasse 190, 8057 Zurich, Switzerland}

\author{Steffen Hagstotz}
\affiliation{Universitäts-Sternwarte, Fakultät für Physik, Ludwig-Maximilians Universität München, 
Scheinerstraße 1, D-81679 München, Germany and\\
Excellence Cluster ORIGINS, Boltzmannstraße 2, D-85748 Garching, Germany}

\begin{abstract}
We develop an analytical framework to predict the one-point probability distribution function (PDF) of dispersion measures (DMs) for fast radio bursts (FRBs) within the baryonification (BFC) model. BFC provides a computationally efficient alternative to expensive hydrodynamical simulations for modelling baryonic effects on cosmological scales. By applying the halo mass function and halo bias, we convolve contributions from individual halos across a range of masses and redshifts to derive the large-scale structure contribution to the DM PDF. We validate our analytical predictions against consistency-check simulations and compare them with the IllustrisTNG hydrodynamical simulation over the redshift range $ z = 0$ to $z = 5$, demonstrating excellent agreement. We demonstrate that our model produces consistent results when fitting gas profiles and predicting the PDF, and vice versa.
We show that the BFC parameters controlling the gas profile, particularly the halo mass scale ($M_\mathrm{c}$), mass-dependent slope ($\mu$), and outer truncation ($\delta$), are the primary drivers of the PDF shape. Additionally, we investigate the validity of the log-normal approximation commonly used for DM distributions, finding that it provides a sufficient description for a few hundred FRBs. Our work provides a self-consistent model that links gas density profiles to integrated DM statistics, enabling future constraints on baryonic feedback processes from FRB observations.

\end{abstract}

\maketitle
\section{Introduction}
Fast radio bursts (FRBs) are brief yet extremely bright transients, lasting only a few milliseconds and spanning frequencies from roughly 100 MHz to several GHz. As the initial pulse travels through the ionised gas in both the \review{host galaxy and the large-scale structure (LSS)\footnote{In the literature this contribution is often called the inter-galactic medium (IGM) contribution.}}, it becomes dispersed: lower-frequency ($\nu$) components arrive later, producing a delay that scales as $\Delta t(\nu) \propto \Delta(1/\nu^{2})$. The proportionality constant, known as the dispersion measure (DM), corresponds to the total column density of free electrons along the line of sight (LOS) to the FRB \citep[see e.g.][]{2007Sci...318..777L, thornton_population_2013, petroff_real-time_2015, connor_non-cosmological_2016, champion_five_2016, chatterjee_direct_2017}.

Although the mechanism responsible for FRB radio emission remains uncertain, their isotropic sky distribution and typically large observed DMs strongly suggest that most events originate outside the Milky Way \citep[though a subset may be Galactic; see e.g.][]{andersen_bright_2020}. 
This extragalactic origin requires modelling the DM in a cosmological setting, but also enables FRBs to be a powerful cosmological probe. One refers to FRBs with an associated host as localised FRBs. The DM-redshift relation of those FRBs has been used to constrain the mean baryon density in the Universe \citep{macquart_census_2020}, the expansion rate \citep{hagstotz_new_2022,james_measurement_2022}, or fundamental physics such as the equivalence principle \citep{2023MNRAS.523.6264R}. A key aspect of these types of analyses is that they require the probability distribution function (PDF) of the DM. Early analyses of hydrodynamical cosmological simulations suggest that the DM induced by the electrons distributed in the LSS produce a PDF well described by a log-normal distribution \citep{2021ApJ...906...49Z,2024A&A...683A..71W}. Given this simple functional form, the PDF can be fully specified by calculating the free-electron variance. It has therefore been realised that the uncertainty in the DM-redshift relation contains valuable information about the one-point statistical properties of the gas distribution \citep[e.g.][]{mcquinn_locating_2014,reischke_cosmological_2023,medlock_probing_2024,2025ApJ...989...81S,2025ApJ...983...46M}. 
This is crucial as cosmological surveys such as Euclid \citep{Mellier2025} or Rubin Observatory Legacy Survey
of Space and Time \citep[LSST,][]{lsst_science_collaboration_lsst_2009} aspire to map the LSS deep into the non-linear regime by measuring the shapes and positions of billions of galaxies.
The distribution of baryonic matter is influenced by feedback processes from Active Galactic Nuclei, Supernovae and others. In turn, these feedback mechanisms also influence the overall matter distribution in the Universe and, consequently, impact cosmological observables.
It is on these non-linear scales that understanding the distribution of baryons has become paramount \citep[e.g.][for an overview]{chisari_modelling_2019}.
\citet{2025arXiv250717742R,2025NatAs.tmp..131C} assumed a log-normal distribution for the DM-redshift relation and put constraints on baryonic feedback using $\sim10^2$ FRBs. 

Recently, \citet{2025arXiv250707090K} remeasured the PDF in the IllustrisTNG simulation, finding considerable disagreements with previous results due to improved line-of-sight integration and interpolation. In particular, they found that the PDF is less well described by a log-normal distribution at low redshifts.
\review{A general question thus arises: how can the multi-phase baryon distribution in the LSS be modelled efficiently?}
While conducting large suites of hydrodynamical simulations with cosmologically relevant volumes would be ideal for studying baryonic feedback, such calculations are computationally prohibitive in a cosmological analysis. As a result, several alternative methods have been developed to capture baryonic effects on the matter distribution, including halo-model-based approaches as well as emulating hydrodynamical simulations\citep{McCarthy_BAHAMAS_2017,mead_hydrodynamical_2020,troster_joint_2022}. Alternatively, one can paint new haloes onto existing 
$N$-body simulations \citep{Osato_Nagai_2023}.
A complementary approach is offered by the so-called baryonification method \citep[first introduced in][]{schneider_new_2015}, which relies on modifications to
$N$-body simulations, using empirical functions to describe the gas, stellar, and dark matter profiles surrounding galaxy groups and clusters. The method can be calibrated against observations or hydrodynamical simulations, offering a computationally efficient means of modelling cosmological probes. In recent years, several models and implementations based on baryonification have been developed \citep[e.g.][]{schneider_quantifying_2019,Giri2021_BCemu,arico_bacco_2021,Anbajagane_bfc_2024}. Baryonification has been shown to reproduce the matter power spectra measured in hydrodynamical simulations across a range of feedback scenarios and observables: the suppression of the matter power spectrum \citep{Giri2021_BCemu}, higher order statistics\citep{arico_simultaneous_2021,2022PhRvD.105h3518F,2023MNRAS.524.5212Z,2023MNRAS.527.1124L,2025arXiv250618974B,2025JCAP...09..073Z}, gas fractions \citep{Schneider:2019xpf,Grandis2023}, or the kinematic Sunyaev-Zel'Dovich (kSZ) effect \citep{2022MNRAS.514.3802S}. 
The most recent version was presented in \citet{2025arXiv250707892S} and is dubbed the baryonification (BFC) model. In \citet{2025arXiv250707991K} it was shown that BFC can consistently reproduce X-ray and kSZ observations.

 In this work, we apply BFC to FRBs to link their observed DM to cosmological observables.
An FRB signal propagates through three distinct environments before reaching the observer. It first travels through its host galaxy, then passes through the IGM, and finally traverses the Milky Way. Although \citet{2025OJAp....8E.127R} studied the host-galaxy contribution to the DM in detail, a comprehensive model for the \review{LSS} component is still lacking. Similar to the host galaxy, the statistical properties of the IGM’s DM contribution depend strongly on baryonic feedback processes. We use the BFC model to develop an analytical model of the IGM contribution to the DM PDF of FRBs. In developing this model, we closely follow the approach of \citet{Thiele_2018}, who performed a similar calculation for the thermal Sunyaev–Zel’dovich effect. A related calculation of the DM one-point PDF was presented by \citet{mcquinn_locating_2014}; however, their analysis did not incorporate BFC and did not attempt accurate comparisons to hydrodynamical simulations. Combining the model presented in this paper with the host galaxy \citep[which was tackled in][]{2025OJAp....8E.127R} and the Milky Way contributions \citep[see e.g.][]{2017ApJ...835...29Y,2024RNAAS...8...17O} will allow us to study FRBs consistently in a cosmological context

This paper is organised as follows. In Section \ref{sec:Methodology}, we provide a brief introduction to the DM of FRBs, introduce the BFC model and describe the methods used in our calculations. In Section \ref{sec:Simulation}, we present the two simulations employed in this work, followed by Section \ref{sec:Results}, where we discuss our results. Finally, in Section \ref{sec:Conclusion}, we conclude by summarising the main findings of this study.
Throughout this paper, we define the virial radius ($r_{200}$) as the radius within which the mean density is 200 times the critical density of the Universe. The corresponding virial mass ($M_{200}$) is the total mass enclosed within this virial radius. To maintain consistency with the IllustrisTNG simulation, we adopt the Planck 2015 cosmological parameters \citep{2016A&A...594A..13P}: baryon density $\Omega_\mathrm{b} = 0.0486$, matter density $\Omega_\mathrm{m} = 0.3089$, scalar spectral index $n_\mathrm{s} = 0.9667$, root-mean-square fluctuation amplitude $\sigma_8 = 0.8159$, and the dimensionless Hubble constant $h = 0.6774$.

\begingroup
\renewcommand*{\arraystretch}{1.5}
\begin{table*}
    \centering
    \begin{tabular}{c c c c c c l}
        \hline\hline
         & \textbf{Parameter} & \textbf{Lower limit} & \textbf{Upper limit} & \textbf{Fiducial} & \textbf{Eq.} & \multicolumn{1}{c}{\textbf{Description}} \\
        \hline
       \multirow{8}{*}{\rotatebox[origin=c]{90}{Varied}}&  $\eta$ & 0 & 0.5 & 0.017 & \ref{eq:fstar} & Defines slope of the total stellar-to-halo mass fraction ($f_\mathrm{star}$ ).\\
        
       & $\mathrm{d}\eta$ & 0 & 0.5 & 0.229 &\ref{eq:fcga} & Defines slope of the central stellar-to-halo mass fraction ($f_\mathrm{cga}$ ).\\
        
       &$N_\mathrm{star}$ & 0 & 0.05 &  0.0074 & \ref{eq:fstar} & Fixes normalisation of $f_\mathrm{star}$ and $f_\mathrm{cga}$ .\\
        
      &  $c_\mathrm{iga}$ & 0 & 1 & 0.0093 & \ref{eq:figa} & Sets normalisation of the inner (cold) gas fraction.\\
        
      &  $\theta_\mathrm{co}$ & 0 & 0.5 & 0.231 & \ref{eq:rhogas} & Determines core size of the gas density profile ($\rho_\mathrm{hga}$). \\
        
      &  $\log_{10} M_\mathrm{c}$ & 11 & 15 & 12.86& \ref{eq:beta} & Sets halo mass scale at which the slope of $\rho_\mathrm{hga}$ is exactly -3/2.\\
        
     &   $\mu$ & 0 & 2 & 0.721 &\ref{eq:beta} & Determines how fast the slope of $\rho_\mathrm{hga}$ changes with halo mass.\\
        
     &   $\delta$ & 4 & 8 & 5.47 & \ref{eq:rhogas} & Defines slope of the outer truncation of $\rho_\mathrm{hga}$ . \\ \hline 
     \multirow{4}{*}{\rotatebox[origin=c]{90}{Fixed}}&   $\alpha$ &\multicolumn{3}{c}{{1}}&\ref{eq:rhogas}& Core transition of the gas profile \\
     &   $\gamma$ &\multicolumn{3}{c}{{1.5}}&\ref{eq:rhogas}& Truncation radius transition of the gas profile \\
    &    $M_\mathrm{star}$ &\multicolumn{3}{c}{$2.5 \times 10^{11} \; h^{-1} M_\odot$} & \ref{eq:fstar} & Reference mass for the stellar profile slope \\
      &  $\zeta$ &\multicolumn{3}{c}{{1.376}}&\ref{eq:fstar}& Low-mass slope of the central galaxy profile \\ \hline
        \end{tabular}
        \vspace{.1cm}
    \caption{All BFC parameters and their respective flat prior limits, as well as the fiducial values obtained from fitting the gas profiles of Illustris TNG directly (compare \Cref{fig:fit_to_profile}). If only a single number for the upper and lower limits is given, the parameter is fixed. The last column gives a brief physical interpretation of the parameter and a reference to the equation in which it first occurs.}
    \label{tab:priorMCMC}
\end{table*}
\endgroup

\vspace{.3cm}
\section{Methodology}
\label{sec:Methodology}
In this section, we first introduce the basic properties of FRBs and their associated DM in Section \ref{sec:frb_general}. In Section \ref{sec:bcm}, we introduce BFC and all relevant profiles used. We calculate the one-point PDF of the DM of the LSS in Section \ref{sec:pdf_methodology}. In Section \ref{sec:comp_other}, we compare our approach to other works. Lastly, in Section \ref{sec:emulator}, we describe an emulator used to speed up analytical calculations.
All cosmological quantities are calculated using the Core Cosmology Library \citep[\texttt{CCL}][]{chisari_core_2019}. Linear matter power spectra are computed with \texttt{CAMB} \citep{lewis_efficient_2000}. Furthermore, we use some functionality of \texttt{astropy} \citep{astropy_collaboration_astropy_2022}.

\subsection{Dispersion Measure of Fast Radio Bursts}
\label{sec:frb_general}
One of the key characteristics of FRBs is their DM. As a radio signal from an FRB at direction $\textbf{n}$ and redshift $z$ travels through the ionised gas between the source and the observer, each frequency travels at a different effective speed, leading to delays quantified by the following relation
\begin{equation}
\Delta t \propto  \; \Delta(1/\nu^{2}).
\end{equation}
The proportionality constant is the DM and is given by the integral of the comoving electron density $n_\mathrm{e}$ along the LOS:
\begin{equation}
\label{eq:DM}
\mathrm{DM}(z,\mathbf{n}) = \int_{0}^{z} n_\mathrm{e}(z^\prime,\mathbf{n})(1+z^\prime)\dfrac{c\mathrm{d}z^\prime}{H(z^\prime)},
\end{equation}
where $c$ is the speed of light and $H(z)$ is the Hubble function.

It is convenient to split the DM into different contributions:
\begin{equation}
\mathrm{DM_{total}}(z,\mathbf{n}) = \mathrm{DM_{MW}}(\mathbf{n}) + \mathrm{DM_{LSS}}(z,\mathbf{n}) + \mathrm{DM_{host}}(z),
\end{equation}
where $\mathrm{DM_{MW}}$, $\mathrm{DM_{LSS}}$, and $\mathrm{DM_{host}}$ are the DMs of the Milky Way, the LSS, and the host galaxy, respectively. The redshift $z$ and positional, $\mathbf{n}$, dependencies are written out explicitly for all components. The Milky Way component does not vary with redshift, and the host term is unaffected by the host galaxy’s spatial location. The former can be measured from galactic Pulsar dispersion measures \citep[e.g.][]{2020ApJ...897..124O} and has been modelled in \citep{2024RNAAS...8...17O} using a simplified model introduced in \citet{2002astro.ph..7156C}. Another work \citep{2017ApJ...835...29Y} introduced a different model for the interstellar medium contribution of the Milky Way; most cosmological results, however, do not depend on this specific choice as long as the galactic disk is avoided \citep[see e.g.][]{2023MNRAS.523.6264R}. $\mathrm{DM_{host}}$ was modelled in \citet{2025OJAp....8E.127R} using the BFC model and was shown to match the output of hydrodynamical simulations \citep{theis_galaxy_2024,medlock_probing_2024}.
Here, we will put forward an equivalent model for $\mathrm{DM_{LSS}}$ based on the same BFC model.

\subsection{Baryonification Model}
\label{sec:bcm}
BFC was proposed and subsequently improved in a series of papers, see e.g. \citet{schneider_new_2015,2025arXiv250707892S} for the initial proposal and the latest implementation. The goal is to model baryonic effects by displacing particles in dark matter-only (DMO) simulations. Since DMO simulations are less complex and computationally intensive than hydrodynamical simulations, BFC offers a faster, more efficient approach to studying the baryon distribution. Crucially, BFC is also more interpretable due to a set of flexible, physically motivated parameters that allow it to describe vastly different feedback scenarios. \Cref{tab:priorMCMC} summarises those parameters.

BFC displaces dark matter and baryonic particles differently; in this work, we focus on the baryonic component. For the baryonic component in BFC, the initial density profile is taken directly from the DMO simulation. It is described as the sum of a truncated NFW profile \citep{navarro_universal_1997, baltz_analytic_2009} and a two-halo term:
\begin{equation}
    \rho_{\mathrm{bar,i}}(r) = f_\mathrm{bar}\left[ \rho_{\mathrm{NFW}}(r) + \rho_\mathrm{2h}(r) \right],
\end{equation}
where $f_\mathrm{bar}= \Omega_\mathrm{b}/\Omega_\mathrm{m}$ is the cosmic baryon fraction. The two-halo term is defined as:
\begin{equation}
    \rho_\mathrm{2h}(r) = \left[ 1+b(\nu) \xi_\mathrm{lin}(r) \right] \Omega_\mathrm{b} \rho_\mathrm{crit}\; ,
\end{equation}
with $b(\nu)$ the halo bias \citep[specifically we use the one presented in][]{2008ApJ...688..709T} and $\xi_\mathrm{lin}(r)$ the linear matter correlation function.

To obtain the final baryonic density profile, the initial NFW component is replaced by the sum of a gas density profile and a stellar density profile:
\begin{equation}
    \rho_\mathrm{bar,f}(r) = \rho_\mathrm{gas}(r) + \rho_\mathrm{star}(r) + f_\mathrm{bar}\rho_\mathrm{2h}(r) .
\end{equation}
Each of these profiles has its own analytical form and depends on a distinct set of BFC parameters, which must be constrained using observational data.

The enclosed mass of each component can be obtained by integrating the density profile.
\begin{equation}
M_\chi(R) = 4\pi \int_{0}^{R} r^2 \rho_{\chi} \,\mathrm{d}r = f_\chi Y_\chi(R),
\end{equation}  
where $f_\chi$ is the abundance of components $\chi$ and $Y_\chi$ is normalised such that 
\begin{equation}
Y_\chi (\infty) = M_\mathrm{tot}.
\end{equation} 
Given this, the total mass of each species is purely determined by its abundance ($f_\chi$).

We then introduce a displacement function, $d(r)$, which maps particle positions in the DMO simulation to those that reproduce the desired baryonic density profile. The mapping is defined through the inverted cumulative mass profile:
\begin{align}
r_\mathrm{bar} &= \mathrm{inv}[M_\mathrm{bar}(r)]\\
d(r) &= r_\mathrm{bar,f} - r_\mathrm{bar,i},
\end{align}
where $\mathrm{inv}[M_\mathrm{bar}(r)]$ denotes the inverse of the cumulative mass profile. This allows us to displace particles in DMO simulations by  
\begin{equation}
r_\mathrm{bar,f} = r_\mathrm{bar,i} + d(r)\,,
\end{equation}
in order to obtain the baryon density profile from the DMO profile.

The gas profile in the BFC model consists of two components: a hot gas component (HGA) and a cold gas component (IGA). We assume that the HGA is fully ionised, while the IGA is mostly neutral and therefore does not directly affect our results. Consequently, our discussion will focus on the HGA. 
The analytic form of the hot gas density profile is modelled as a cored double power law:
\begin{equation}
    \label{eq:rhogas}
    \rho_{\mathrm{hga}}(r) \propto \dfrac{f_\mathrm{bar}-f_\mathrm{star}(M_{200})-f_\mathrm{iga}(M_{200})}{\left[ 1+\left(\frac{r}{r_\mathrm{c}}\right)^\alpha \right]^{\beta/\alpha} \left[ 1+\left(\frac{r}{r_\mathrm{t}}\right)^\gamma \right]^{\delta/\gamma}},
\end{equation}
where $M_{200}$ is the halo mass, $r_\mathrm{c} = \theta_\mathrm{c} r_{200}$ denotes the core radius, and $r_t = \epsilon r_{200}$ specifies the truncation radius. Throughout this work, the truncation exponent is fixed to $\gamma = 3/2$.

The slope $\beta$ of the gas profile varies with halo mass. When accounting for AGN feedback, lower-mass haloes typically exhibit shallower gas density profiles, i.e. lower $\beta$. This behaviour is captured by the following parametrisation:
\begin{equation}
    \label{eq:beta}
    \beta(M_{200}) = \dfrac{3(M_{200}/M_\mathrm{c})^\mu}{1+(M_{200}/M_\mathrm{c})^\mu}\;.
\end{equation}
The stellar profile is assumed to follow the dark matter distribution:
\begin{equation}
\rho_{\rm sga}(r)=\frac{f_{\rm sga}}{4\pi r^2}\frac{\mathrm{d}}{\mathrm{d}r}M_{\rm nfw}(\xi r)\;,
\end{equation}
where back-reaction was included. Note that there is a separate profile for the central galaxy. We parametrise the stellar mass fraction as:
\begin{equation}
    \label{eq:fstar}
    f_\mathrm{star}(M_{200}) = \dfrac{N_\mathrm{star}}{\left( \frac{M_{200}}{M_\mathrm{star}}\right)^{-\zeta} +\left( \frac{M_{200}}{M_\mathrm{star}}\right)^{\eta}} \; .
\end{equation}
Lastly, the cold gas profile is 
\begin{equation}\label{igaprofile}
\rho_{\rm iga}(r) = \frac{\rho_{\rm iga,0}}{r^3}\exp\left(-r/r_{200}\right)\;,
\end{equation}
with the corresponding fraction defined as
\begin{equation}
    \label{eq:figa}
    f_\mathrm{iga}(M_{200}) = c_\mathrm{iga} f_\mathrm{cga}(M_{200}),
\end{equation}
with $f_\mathrm{cga}(M)$ given by
\begin{equation}
    \label{eq:fcga}
    f_\mathrm{cga}(M_{200}) = \dfrac{N_\mathrm{star}}{\left( \frac{M_{200}}{M_\mathrm{star}}\right)^{-\zeta} +\left( \frac{M_{200}}{M_\mathrm{star}}\right)^{\eta+ \mathrm{d}\eta}} \; ,
\end{equation}
where $\zeta = 1.376$ and $M_\mathrm{star}= 2.5 \times 10^{11} \; h^{-1} M_\odot$.
All parameters, except for $\epsilon$, which depends on the choice of the NFW profile, are free parameters and influence the shape of the profile. We will examine the impact of these parameters on the final PDF later. \Cref{tab:priorMCMC} summarises the parameters used here, together with a quick reference for each parameter to its respective occurrence in the model.

\subsection{DM one-point PDF of LSS}
\label{sec:pdf_methodology}
In this section, we explain how to calculate the one-point PDF for the LSS. First, we begin with the DM profile of a single halo, then closely follow the methodology of \citet{Thiele_2018}, who performed a similar calculation for the thermal Sunyaev-Zel'dovich effect. For completeness, we rederive all relevant expressions presented in  \citet{Thiele_2018}. Throughout this section, for notational convenience, we denote all halo masses by $M$ rather than $M_{200}$.

\begin{figure}
    \centering
    \includegraphics[width=.95\linewidth]{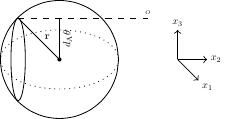}
    \caption{Geometry of the projected DM as illustrated in \Cref{eq:DM_profile}. Here $d_\mathrm{A}$ is the angular diameter distance at the redshift of the halo. The projection is carried out along the $x_2$-axis towards the observer $O$.}
    \label{fig:Halo_scheme}
\end{figure}

\citet{2025OJAp....8E.127R} presented an analytical model for the DM of FRBs within their host halos based on BFC. 
Here, the situation is slightly different. The geometry is shown in \Cref{fig:Halo_scheme}. We want to know the contribution to the DM of a halo of mass $M$ and at redshift $z$  which has been traversed by an FRB at angle $\theta$ from the centre.
Suppose $\mathrm{DM}^{(3)}_{M,z}(x_1,x_2,x_3)$ is the DM of an FRB at position $\boldsymbol{r}^\mathrm{T} =(x_1,x_2,x_3)$ relative to the halo centre, where we project along the $x_3$ axis without loss of generality due to spherical symmetry. As a result, the DM profile depends only on the radial distance from its centre. We can write the two-dimensional DM profile, $\mathrm{DM}(M,z,\theta)$, of the isolated halo under consideration as:
\begin{equation}
    \label{eq:DM_profile}
    \mathrm{DM}(M,z,\theta) = \mathrm{DM}^{(3)}_{M,z}(0,-\sqrt{r^2-(d_\mathrm{A} \theta)^2},d_\mathrm{A}\theta),
\end{equation}
where $d_\mathrm{A} =d_\mathrm{A}(z) $ is the comoving angular diameter distance to the halo centre, $\theta$ is the angular separation on the sky, $\mathrm{DM}^{(3)}_{M,z}$ is the three-dimensional DM profile within the halo, and $r$ is the halo radius. The DM of such a halo can also be written as:
\begin{equation}
\label{eq:DM_Halo}
    \mathrm{DM}(M,z,\theta) =  \chi_\mathrm{e}(1+z)^2\int \mathrm{d}x_2\; \frac{\rho_\mathrm{hga}[r(x_2,\theta,z), M,z]}{\mu_\mathrm{e}m_\mathrm{p}}.
\end{equation}
 Note that all quantities in this equation are comoving.
Furthermore, we have the proton mass $m_\mathrm{p}$, the mean molecular weight per electron $\mu_\mathrm{e} = 1.17$ and the electron fraction $\chi_\mathrm{e}$ given by:
\begin{align}
\label{eq:chi_e}
    \chi^{\mathstrut}_\mathrm{e} &= Y^{\mathstrut}_\mathrm{H} + \frac{1}{2} Y^{\mathstrut}_\mathrm{He} \\
    & \approx 1 - \frac{1}{2} {Y}^{\mathstrut}_\mathrm{He} \, .
\end{align}
This is calculated from the primordial hydrogen and helium abundances $Y_\mathrm{H}$ and $Y_\mathrm{He}$. We assume $Y_\mathrm{H} \approx 1 - Y_\mathrm{He}$ and $Y_\mathrm{He} = 0.245$  \citep{2015JCAP...07..011A,planck_collaboration_planck_2020}.

To calculate the one-point PDF of the FRB DMs, we proceed as follows: We first consider a small volume element and a mass bin in which halo overlap can be neglected. Then, we compute the PDF within this region and afterwards convolve all halo mass bins along the LOS to account for overlapping halos. Finally, to incorporate halo clustering, we compute the one-point PDF for a fixed linear field $\delta_\mathrm{lin}(\textbf{n},z)$ and average over Gaussian realisations of the field $\delta_\mathrm{lin}$.
As per usual, it is advantageous to work in Fourier space due to the convolution theorem.
Throughout, we will use the following Fourier convention, so that the Fourier transform of the one-point PDF is:
\begin{equation}
    \tilde{P}(\lambda) = \int \mathrm{d}\mathrm{DM}\;P(\mathrm{DM}) e^{\mathrm{i}\lambda \mathrm{DM}}. 
\end{equation}
$\tilde{P}(\lambda)$ can then be written as 
\begin{equation}
\label{eq:expectaion_value}
    \tilde{P}(\lambda) = \langle \exp{\left[\mathrm{i}\lambda \mathrm{DM}(\textbf{n})\right]}\rangle,
\end{equation}
where $\mathrm{DM}(\textbf{n})$ is the DM at a fixed sight-line $\textbf{n}$. This is simply the characteristic function of the random variable DM.
The expectation value in \Cref{eq:expectaion_value} can be calculated by averaging over random halo placements. 
The number of halos in the volume element $\mathrm{d}V$ with masses between $M$ and $M+\mathrm{d}M$ is given by
\begin{equation}
    \mathrm{d}N = \dfrac{\mathrm{d}n(M,z)}{\mathrm{d}M}\mathrm{d}V \mathrm{d}M,
\end{equation}
where ${\mathrm{d}n(M,z)}/{\mathrm{d}M}$ is the halo mass function. 
The angular halo number density in these narrow bins is
\begin{equation}
    \dfrac{\mathrm{d}n}{\mathrm{d}\Omega} = \dfrac{\chi^2(z)}{H(z)} \dfrac{\mathrm{d}n(M,z)}{\mathrm{d}M} \mathrm{d}M\mathrm{d}z ,
\end{equation}
where $\chi(z)$ is the comoving distance to redshift $z$.
If we choose $\mathrm{d}z$ and $\mathrm{d}M$ to be sufficiently small, then overlaps between halos can be neglected. \review{For numerical purposes, we assume that each halo has a finite angular radius $\theta_{\mathrm{max}}$, beyond which its contribution to the DM can be neglected. Typically, we use $\theta_{\mathrm{max}}$ to be five times the virial radius. We discuss this choice in more detail in 
Appendix \ref{app:rmax}. Contributions beyond that will be dominated entirely by the two-halo term and will not change the shape of the PDF but only its mean.}

With the assumptions discussed previously, \Cref{eq:expectaion_value} can be replaced with an integral over the angular DM function as described in \Cref{eq:DM_profile}, i.e.
\begin{equation}
\label{eq:p_lambda}
     \tilde{P}(\lambda) = \left(1-\dfrac{\mathrm{d}n}{\mathrm{d}\Omega} \pi \theta_{\mathrm{max}}^2 \right) + \dfrac{\mathrm{d}n}{\mathrm{d}\Omega} \int_0^{\theta_{\mathrm{max}}} \mathrm{d}\theta 2\pi \theta e^{\mathrm{i}\lambda \mathrm{DM}(\theta)}.
\end{equation}
The first term represents the probability that no halo overlaps with the sight-line $\textbf{n}$, while the second term integrates over the overlapping locations.

\Cref{eq:p_lambda} can be rewritten in the following compact form
\begin{equation}
\label{eq:p_lambda_narrow_exact}
    \tilde{P}(\lambda) = 1+ \dfrac{\chi^2(z)}{H(z)} \dfrac{\mathrm{d}n(M,z)}{\mathrm{d}M} \tilde{Y}(M,z,\lambda) \mathrm{d}M\mathrm{d}z,
\end{equation}
where 
\begin{equation}
    \tilde{Y}(M,z,\lambda) = \int \mathrm{d}\theta 2 \pi \theta \left( e^{\mathrm{i}\lambda \mathrm{DM}(M,z,\theta)} -1 \right).
\end{equation}

At this point, we can calculate the one-point PDF considering all masses and redshifts. For now, we assume that we can neglect halo clustering; thus, the contribution to the DM from each halo bin is independent. For $N$ bins $(M_1,z_1) \cdots (M_N,z_N)$, the complete PDF can be calculated by convolution.

For a differential bin ($\mathrm{d}M$ $\mathrm{d}z$) \Cref{eq:p_lambda_narrow_exact} is equivalent to 
\begin{equation}
    \tilde{P}(\lambda) = \exp \left( \dfrac{\chi^2(z)}{H(z)} \dfrac{\mathrm{d}n(M,z)}{\mathrm{d}M} \tilde{Y}(M,z,\lambda) \mathrm{d}M\mathrm{d}z \right).
\end{equation}
Since the convolution becomes multiplication in Fourier space, we will have:
\begin{align}
    \tilde{P}(\lambda) &= \lim_{N \rightarrow \infty} \prod_{i=1}^{N} \tilde{P}_{M_i,z_i}(\lambda) \nonumber \\
    &= \exp{ \int  \dfrac{\chi^2(z)}{H(z)} \dfrac{\mathrm{d}n(M,z)}{\mathrm{d}M} \tilde{Y}(M,z,\lambda) \mathrm{d}M \mathrm{d}z} .\label{eq:PDF_without_clustering}
\end{align}
 \Cref{eq:PDF_without_clustering} is the final result for the one-point PDF of the DM in Fourier space, neglecting halo clustering.
Note that the PDF derived in this way does not capture the mean contribution to the DM, as we would have to include the two-halo term up to $r\to\infty$, which is numerically difficult as it provides a constant term to the Fourier transform. We discuss this in more detail in Section \ref{sec:comp_cons_sim}. The model, however, stays fully predictive as the mean (see Equation \ref{eq:DM_LSS_avg}) can be computed directly from the cosmological parameters.

We now incorporate halo clustering into the PDF. We denote the unclustered PDF (Equation \ref{eq:PDF_without_clustering}) and clustered PDF by $\tilde{P}_\mathrm{u}(\lambda)$ and $\tilde{P}_\mathrm{cl}(\lambda)$, respectively.
The derivation involves two main steps. First, we calculate the one-point PDF $\tilde{P}_\delta(\lambda,\textbf{n})$ for a fixed realisation of the linear density field $\delta_{\mathrm{lin}}(\textbf{n},z)$. Next, we average over all realisations of the Gaussian field $\delta_{\mathrm{lin}}$ to obtain $\tilde{P}_\mathrm{cl}(\lambda)$. (Note that $\tilde{P}_\delta$ depends on $\textbf{n}$ because a specific realisation of $\delta_\mathrm{lin}$ breaks translational invariance, while $\tilde{P}_\mathrm{cl}(\lambda)$ remains translationally invariant, as expected.)

$\tilde{P}_\delta(\lambda,\textbf{n})$ can be calculated by biasing the halo mass function with the halo bias $b(M,z)$ in \Cref{eq:PDF_without_clustering}. This corresponds to a peak–background split: the large-scale linear density field provides a background that modulates the local number of halos, which is captured by the biasing with $\delta_\mathrm{lin}$:
\begin{align}
  \tilde{P}_\delta(\lambda,\textbf{n}) = 
  \exp & \int \mathrm{d}M \mathrm{d}z  \dfrac{\chi^2(z)}{H(z)} \tilde{Y}(M,z,\lambda) \nonumber \\ 
  &\left\{ \dfrac{\mathrm{d}n(M,z)}{\mathrm{d}M}\left[ 1 + b(M,z)\delta_\mathrm{lin}(\textbf{n},z) \right] \right\} .
    \label{eq:first_cluster}  
\end{align}
This expression assumes that the linear density field varies only on scales much larger than the typical size of a halo, such that one can define a sufficiently large region around $\textbf{n}$ over which $\delta_\mathrm{lin}(\textbf{n}+\textbf{n}^\prime,z) \simeq \delta_\mathrm{lin}(\textbf{n},z)$. In this regime, the long-wavelength linear density field acts as a locally constant background that modulates the mean halo abundance.

\begin{figure*}
    \centering
    \begin{subfigure}[b]{0.5\textwidth}
         \centering
         \includegraphics[scale=1.2]{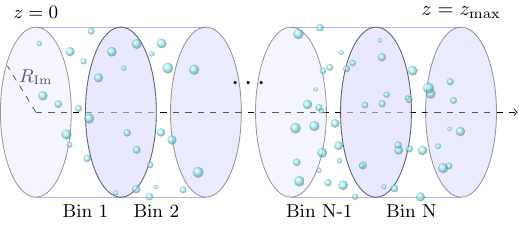}
         \caption{}
         \label{fig:2a}
     \end{subfigure}
     \hfill
     \begin{subfigure}[b]{0.3\textwidth}
         \centering
         \includegraphics[scale=.45]{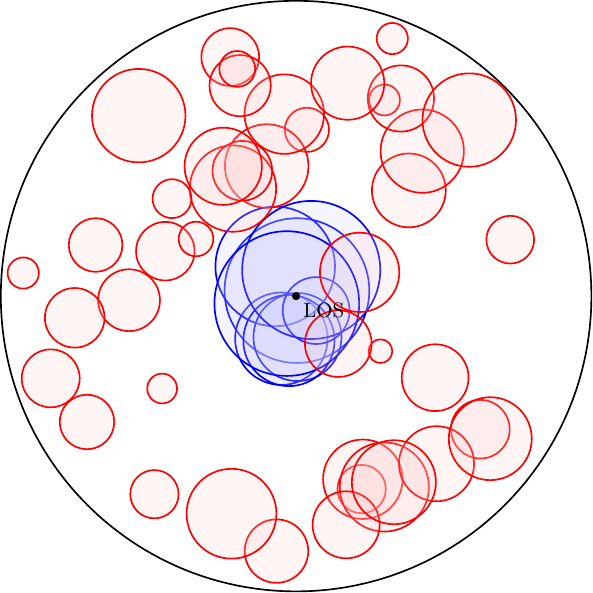}
         \caption{}
         \label{fig:2b}
     \end{subfigure}
     \caption{Figure (a) illustrates an example of the simulation setup, where the volume is divided into equally spaced redshift bins. Figure (b) displays the halos projected onto a plane for an arbitrary bin. Halos that intersect with the LOS are shown in light blue, and only their contributions are included in the analysis. This serves as a consistency-check simulation, as it must reproduce, by definition, the analytical treatment of the unclustered case \Cref{eq:PDF_without_clustering}.}
     \label{fig:simulation}
\end{figure*}

\begin{figure}
    \centering
    \includegraphics[width=0.45\textwidth]{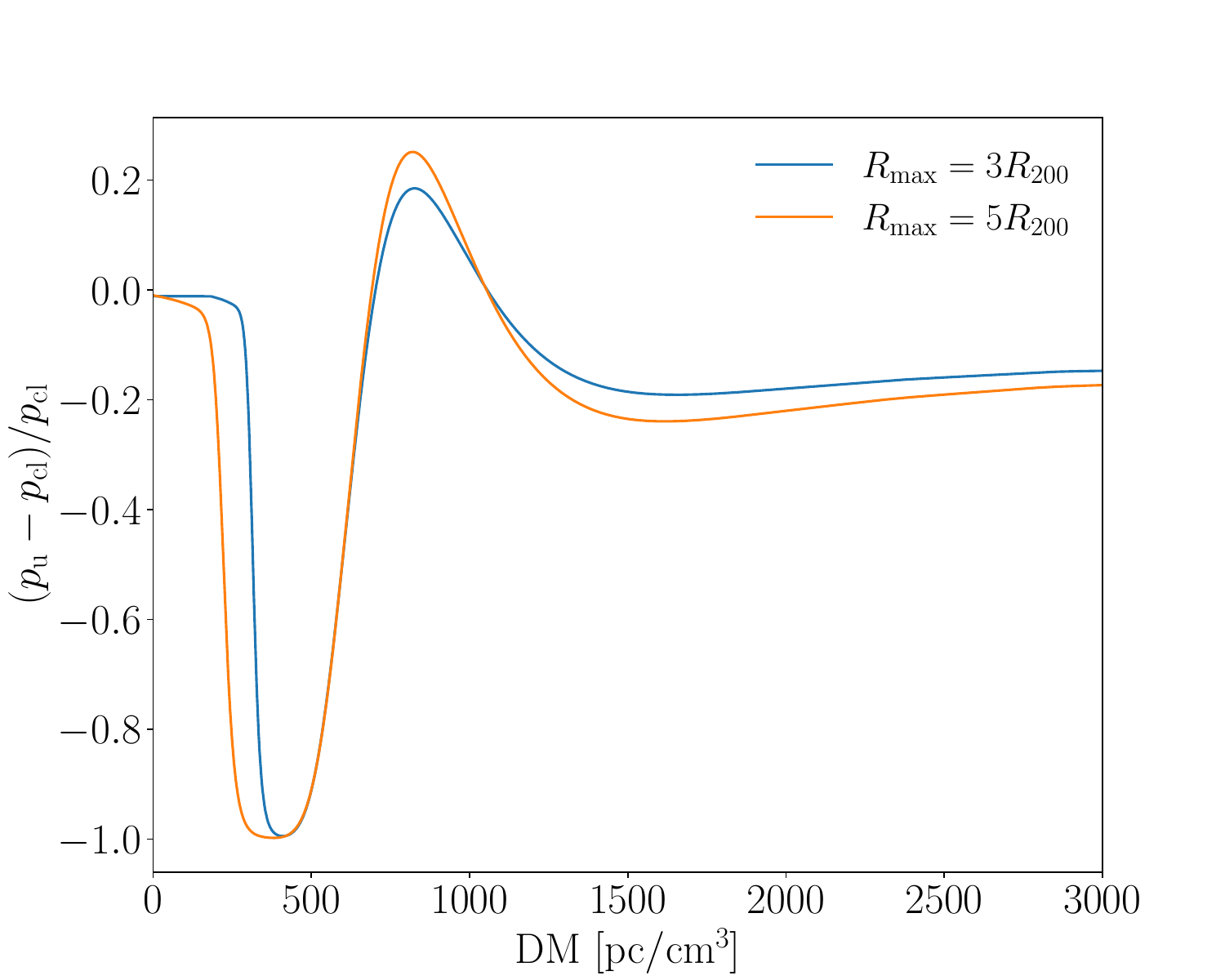}
    \caption{\review{The relative effect of clustering on the DM PDF as described in \Cref{eq:final_analytic}. We show the effect for two different maximum radii out to which we integrate the hot gas profile.}}
    \label{fig:clustering_comp}
\end{figure}

Now we need to average over realisations of the linear density field in \Cref{eq:first_cluster}:
\begin{equation}
    \tilde{P}_{\mathrm{cl}}(\lambda) = \big\langle \tilde{P}_\delta(\lambda,\textbf{n})\big\rangle_{\delta(\textbf{n},z)}. \label{eq:second_cluster}
\end{equation}
By defining two auxiliary functions
\begin{align}
    A(\lambda,\textbf{n}) &\equiv \int \mathrm{d}z \delta_\mathrm{lin}(\textbf{n},z) \alpha(z,\lambda) \label{eq:third_cluster}, \\
    \alpha(z,\lambda) &\equiv \int \mathrm{d}M b(M,z) \dfrac{\chi^2(z)}{H(z)} \dfrac{\mathrm{d}n(M,z)}{\mathrm{d}M} \tilde{Y}(M,z,\lambda),
\end{align}
\Cref{eq:first_cluster} can be written as:
\begin{equation}
    \tilde{P}_\delta(\lambda,\textbf{n}) = \tilde{P}_u(\lambda) e^{A(\lambda,\textbf{n})}.
\end{equation}
By using the identity $\langle \mathrm{e}^x \rangle = \mathrm{e}^{\frac{1}{2}\langle x^2 \rangle}$ for a Gaussian random variable $x$, \Cref{eq:second_cluster} can be rewritten as:
\begin{equation}
    \tilde{P}_{\mathrm{cl}}(\lambda) = \tilde{P}_\mathrm{u} (\lambda) \exp \frac{1}{2} \left\langle A^2(\lambda,\textbf{n})\right\rangle_\delta.
\end{equation}
The quantity $\langle A^2 \rangle$ can be calculated using linear perturbation theory.
In this regime, the density contrast factorises into a time-dependent growth
factor normalised to one at redshift zero and a purely spatial field,
\begin{equation}
\delta_{\mathrm{lin}}(\mathbf{x},z)
= D(z)\,\delta_{\mathrm{lin}}(\mathbf{x}) \; .
\end{equation}
Along the LOS, where $\mathbf{x}=\chi(z)\mathbf{n}$, this relation becomes
\begin{equation}
\delta_{\mathrm{lin}}(\mathbf{n},z)
= D(z)\,\delta_{\mathrm{lin}}\!\left(\chi(z)\mathbf{n}\right) \; .
\end{equation}
Then one can calculate $\langle A^2(\lambda,\textbf{n})\rangle_\delta$ as follows:
\begin{align}
    \langle A^2(\lambda,\textbf{n})\rangle_\delta = \int \mathrm{d}z \mathrm{d}z^\prime D(z)D(z^\prime) \alpha(z,\lambda) \alpha(z^\prime,\lambda) \nonumber \\
    \langle \delta_\mathrm{lin}(\chi(z)\textbf{n})\delta_\mathrm{lin}(\chi(z^\prime)\textbf{n})\rangle .
    \label{eq:A^2_before_limber}
\end{align}
We now evaluate the two-point correlator
by expressing the density field in Fourier space:
\begin{align}
    &\langle \delta_\mathrm{lin}(\chi(z)\textbf{n})\delta_\mathrm{lin}(\chi(z^\prime)\textbf{n})\rangle \nonumber \\
    &=  \int \dfrac{\mathrm{d^3}k}{(2\pi)^3}\dfrac{\mathrm{d^3}k^\prime}{(2\pi)^3} \langle \delta_\mathrm{lin}(\textbf{k})\delta_\mathrm{lin}(\textbf{k}^\prime)\rangle e^{\mathrm{i}\left[\chi(z)\textbf{k}+\chi(z^\prime)\textbf{k}^\prime\right]\cdot\textbf{n}} ,
    \label{eq:delta_to_power_s}
\end{align}
by definition, $\langle \delta_\mathrm{lin}(\textbf{k})\delta_\mathrm{lin}(\textbf{k}^\prime)\rangle = (2\pi)^3 \delta_\mathrm{D}\!\left(\mathbf{k} +\mathbf{k}^\prime\right)P_\mathrm{lin}(k)$, where $P_{\mathrm{lin}}(k)$ denotes the linear matter power spectrum.
Using this relation and performing the integral over $\mathbf{k}'$,
\Cref{eq:delta_to_power_s} can be rewritten as
\begin{align}
    &\langle \delta_\mathrm{lin}(\textbf{n})\delta_\mathrm{lin}(\textbf{n})\rangle \nonumber \\ \nonumber &= \int \dfrac{\mathrm{d^3}k}{(2\pi)^3}\dfrac{\mathrm{d^3}k^\prime}{(2\pi)^3} (2\pi)^3 \delta_\mathrm{D}\!\left(\mathbf{k} +\mathbf{k}^\prime\right)P_\mathrm{lin}(k)  e^{\mathrm{i}\left[\chi(z)\textbf{k}+\chi(z^\prime)\textbf{k}^\prime\right]\cdot\textbf{n}} \\
    &=\int \dfrac{\mathrm{d^3}k}{(2\pi)^3} P_\mathrm{lin}(k) e^{\mathrm{i}\left[\chi(z)-\chi(z^\prime) \right]\textbf{k}\cdot\textbf{n}} .
    \label{eq:before_limber}
\end{align}
At this stage, we decompose the wavevector $\mathbf{k}$ into components
parallel and perpendicular to the LOS, denoted by
$k_\parallel$ and $\mathbf{k}_\perp$, respectively.
Applying the Limber approximation,
\begin{equation}
P_{\mathrm{lin}}\!\left(\sqrt{k_\perp^{\,2}+k_\parallel^{\,2}}\right)
\simeq P_{\mathrm{lin}}(k_\perp),
\end{equation}
\Cref{eq:before_limber} can then be written as
\begin{align}
    &\int \dfrac{\mathrm{d^3}k}{(2\pi)^3} P_\mathrm{lin}(k) e^{\mathrm{i}\left[\chi(z)-\chi(z^\prime) \right]\textbf{k}\cdot\textbf{n}} \nonumber \\
    &= \int \dfrac{\mathrm{d^2 k_\bot}}{(2\pi)^2} P_\mathrm{lin}(k_\bot) \int \dfrac{\mathrm{d}k_\parallel}{2\pi}e^{\mathrm{i}\left[\chi(z)-\chi(z^\prime) \right]k_\parallel} \nonumber \\
    &= \int \dfrac{\mathrm{d^2 k_\bot}}{(2\pi)^2} P_\mathrm{lin}(k_\bot) \delta_\mathrm{D}\!\left[ \chi(z) - \chi(z^\prime) \right] .
    \label{eq:limber}
\end{align}
Substituting \Cref{eq:limber} into \Cref{eq:A^2_before_limber} and using
the identity
\begin{equation}
\delta_\mathrm{D}\!\left[\chi(z)-\chi(z')\right]
= H(z)\,\delta_\mathrm{D}\!\left(z-z'\right),
\end{equation}
we obtain
\begin{equation}
    \langle A^2(\lambda,\textbf{n})\rangle_\delta = \int \mathrm{d}z H(z) D^2(z) \alpha^2(z,\lambda)  \int \dfrac{k\mathrm{d}k}{2\pi} P_\mathrm{lin}(k).
\end{equation}
With all the previous calculations, the final expression for the one-point PDF in Fourier space is given by:
\begin{align}
\tilde{P}_{\mathrm{cl}}(\lambda) = \tilde{P}_\mathrm{u} (\lambda) 
\exp \Bigg( \frac{1}{2} \int \mathrm{d}z\, H(z) D^2(z) \alpha^2(z,\lambda) \nonumber \\
\times \int \frac{k\,\mathrm{d}k}{2\pi} P_\mathrm{lin}(k) \Bigg).
\label{eq:final_analytic}
\end{align}
Unless stated otherwise, we will always use this version of the PDF and not the unclustered one. \review{In \Cref{fig:clustering_comp}, we show the relative effect of clustering on the DM PDF at redshift 0.7. The mean of the PDF is expected to be around 700$\,\mathrm{pc/cm}^{3}$. We thus see that the clustered PDF exceeds the unclustered one in the wings of the distributions. In the core, the situation is reversed, and hence clustering broadens the PDF. The difference in the high DM tail reaches up to 20\%, similar to what was found in \citet{2020PhRvD.102l3545T}. We do not find a strong dependence on redshift. Unless stated otherwise, we use the version including clustering, \Cref{eq:final_analytic}, for all calculations.}

\subsection{The cosmological mean}
It is important to note that the PDF so far considers only electron density fluctuations, $\delta_\mathrm{e}$, and not the mean component. To obtain the observable DM PDF, the mean contribution, which depends solely on cosmological parameters and redshift, must be added to the result. In principle, we would add a two-halo term to the gas profile to make it converge to the cosmic mean on large scales. This, however, is numerically disadvantageous, as it yields a constant in Fourier space and a Dirac delta contribution in DM space, namely the mean. It is therefore much easier to simply compute the mean and add it later. This mean is given by the Macquart relation \citep[see e.g.][]{ioka_cosmic_2003,inoue_probing_2004, deng_cosmological_2014} and can be directly calculated using \Cref{eq:DM}:
\begin{equation}
\label{eq:DM_LSS_avg}
    \langle\mathrm{DM}^{\mathstrut}_\mathrm{LSS}\rangle(z) =\frac{3 \Omega_\mathrm{b} \chi_\mathrm{H}}{8 \uppi G m_\mathrm{p}}\chi_\mathrm{e} \,  \int_0^z \,f^{\mathstrut}_\mathrm{IGM}(z^\prime) \frac{1+z'}{E(z')} \mathrm{d} z' ,
\end{equation}
with the dimensionless baryon density parameter $\Omega_\mathrm{b}$ today, the Hubble radius, $\chi_\mathrm{H} = c/H_0$.
$f_\mathrm{IGM}(z)$ is the gas fraction in the IGM, which can essentially be predicted by BFC as the gas fraction:
\begin{equation}
    f_\mathrm{IGM}(z) = \frac{\int \mathrm{d}M\,f_\mathrm{hga}(M,z) M \frac{\mathrm{d}n}{\mathrm{d}M}(M,z)}{\int \mathrm{d}M\,M \frac{\mathrm{d}n}{\mathrm{d}M}(M,z)}.
\end{equation}

\subsection{Comparison to other works}
\label{sec:comp_other}
As discussed before, \citet{mcquinn_locating_2014} presented an analysis very similar to ours. In the following, we show that the two approaches are equivalent.

The expression for the PDF of the LSS contribution to the FRB DM derived in \citet{mcquinn_locating_2014} is given by:
\begin{equation}
\tilde{P}(\lambda|z_s) = \exp \left[ \int_{0}^{\chi(z_s)} \mathrm{d}\chi \left( A + \frac{B^2 \, \Delta \chi \, \sigma^2_{\Delta\chi}}{2} \right) \right],
\label{eq:mcq1}
\end{equation}
where
\begin{align}
&A = \int \mathrm{d}M \, \mathrm{d^2}R \, a^{-2} \, \frac{\mathrm{d}n(M,z)}{\mathrm{d}M} 
      \left( e^{-i\lambda \Delta_{\mathrm{DM}}(R,M)} - 1 \right), \\
&B = \begin{aligned}[t]
       \int \mathrm{d}M \, \mathrm{d^2}R \, a^{-2} \, \frac{\mathrm{d}n(M,z)}{\mathrm{d}M} \, b(M,z) \\
       \times \left( e^{-\mathrm{i}\lambda \Delta_{\mathrm{DM}}(R,M)} - 1 \right),
     \end{aligned} \\
&\sigma^2_{\Delta \chi} = \int \frac{\mathrm{d^2} k_\perp \, \mathrm{d}k_\parallel}{(2\pi)^3} \, P_\delta(k) \, 
      \mathrm{sinc}^2 \left( \frac{\Delta \chi \, k_\parallel}{2} \right).
\end{align}
Here, $a$ denotes the scale factor, $R$ is the proper impact parameter, and $\Delta_{\mathrm{DM}}(R,M)$ represents the DM profile as a function of radius.

\begin{figure}
    \centering
    \includegraphics[width=\linewidth]{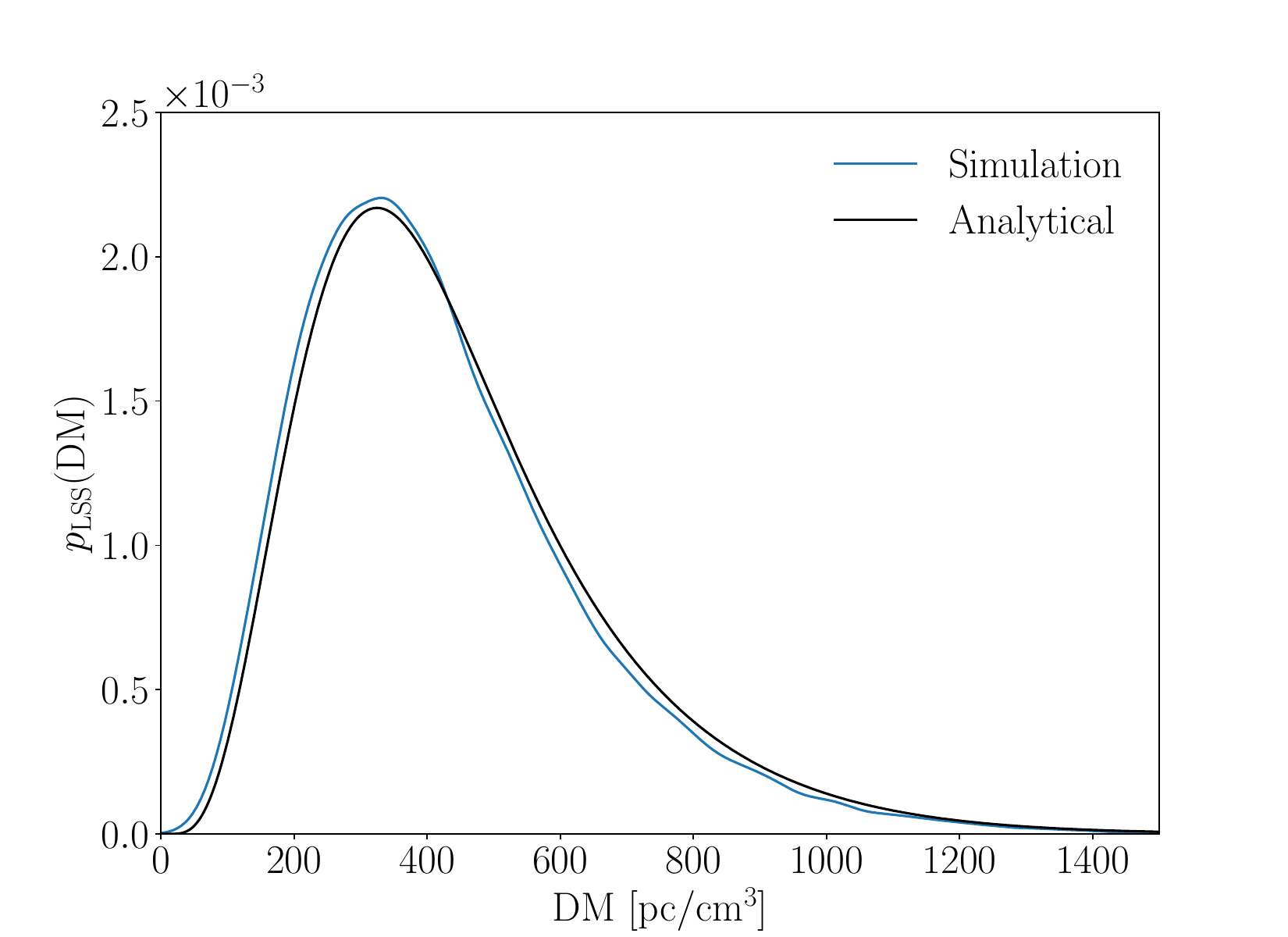}
    \vspace{-.2cm}
    \caption{Comparison between the DM PDF obtained from the \review{consistency-check simulations} (blue) and the analytical result (black) at redshift $z = 1.5$. Both methods use the same BFC parameters and account only for the electron density fluctuations $\delta_\mathrm{e}$, excluding the mean contribution. The shaded region represents the $3\sigma$ confidence interval calculated using the covariance matrix.}
        \vspace{.15cm}
    \label{fig:simulation_comp}
\end{figure}

The first term in \Cref{eq:mcq1} is equivalent to \Cref{eq:PDF_without_clustering}. Noting that $\mathrm{d}z /H(z)\, = \mathrm{d}\chi$, the factor $a^{-2}$ arises from converting the proper impact parameter to the comoving one, and the area element $\mathrm{d^2}R$ corresponds to $\chi^2 \left(2 \pi \theta \,\mathrm{d}\theta \right)$. 

In the second term, the only difference is that we carry out the line-of-sight integral of the matter power spectrum directly, whereas \citet{mcquinn_locating_2014} assumes a plane-parallel approximation. Overall, \Cref{eq:mcq1} is equivalent to \Cref{eq:final_analytic}.

\begin{figure*}
    \centering
    \includegraphics[width=1\linewidth]{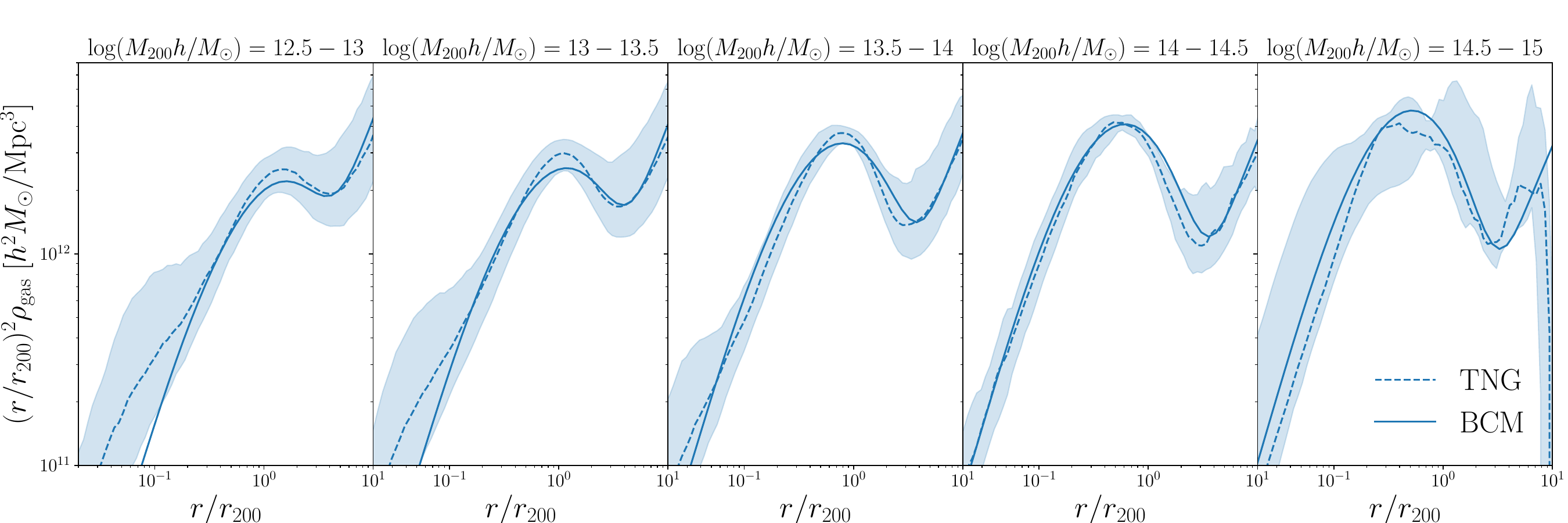}
    \caption{Comparison of the gas density profiles from the TNG simulation and the BFC model. The solid lines represent the BFC profiles, while the dashed lines show the TNG results. The shaded region indicates the 68\% confidence interval of the TNG simulation. All profiles are evaluated in comoving coordinates at redshift 0.5.}
    \label{fig:fit_to_profile}
\end{figure*}

\subsection{Emulator}
\label{sec:emulator}
The pipeline for computing the analytical result presented in \Cref{eq:final_analytic} is computationally expensive. To reduce computation time, we train a machine learning model to emulate the output, enabling rapid evaluations without recalculating the full expression for each parameter set. For this purpose, we use {\sc CosmoPower}, a library designed explicitly for machine-learning-accelerated Bayesian inference \citep[for more details, see][]{2022MNRAS.511.1771S}.

Although {\sc CosmoPower} was initially designed for cosmological power spectrum emulation, its methods are
general and applicable to a wide range of scientific problems. In our work, we use {\sc CosmoPower} to train a model that emulates the PDF. For more details, we refer the reader to Appendix \ref{app:cosmopower}.

\vspace{.3cm}
\section{Simulation and covariance}
\label{sec:Simulation}
In this section, we discuss two types of simulations, of varying complexity, which we employ to evaluate the performance of our analytical model. First, we present a simple consistency-check simulation assuming only random halo placements and analytical profiles. Those simulations are also used to estimate the covariance of the DM PDF.
Second, we use the results from the IllustrisTNG hydrodynamical simulation to demonstrate that our model accurately predicts the DM PDF.

\subsection{Consistency-check simulation}

To validate our analytical results, we perform a Poisson simulation to obtain the DM PDF at various redshifts. \review{We start from a mass function and calculate the number of expected halos within mass and redshift bins $m_i$, $z_j$ respectively. From this number, $\bar N_{m_i,z_j}$, we sample a Poisson random variable $N_{m_i,z_j}$ and randomly place the halos in the corresponding volume of the shell defined by the redshift width. Next, we assign the gas profile to each halo, \Cref{eq:rhogas}, and subsequently project it onto a plane (see \Cref{fig:simulation} for a sketch).  We compute the DM by evaluating the contribution from halos intersecting the LOS. This provides a single DM value. By repeating this procedure for different sightlines, we can generate a histogram of the DM values, which serves as an estimate of the PDF. This corresponds to a single realisation of the consistency-check simulation.}

\subsection{Covariance estimation}
By running multiple realisations of the Poisson simulation, we estimate the covariance of the resulting PDFs as the sample covariance with components:
\begin{equation}
\label{eq:cov}
\textbf{C}_{ij} = \frac{1}{n - 1} \sum_{k=1}^{n} (X_i^{(k)} - \bar{X}_i)(X_j^{(k)} - \bar{X}_j),
\end{equation}
where $n$ is the number of simulation realisations, $X_i^{(k)}$ is the value of the PDF in bin $i$ from the $k$-th realisation and $\bar{X}_i$ is the mean value of the PDF in bin $i$ across all simulations.
Theoretically, this covariance has two contributions: one induced from the LSS directly, i.e. cosmic variance, and a second one from the finite amount of sightlines available, i.e. the number of FRBs.
If the number of sightlines sampled tends to infinity, this corresponds to the cosmic-variance-induced covariance of the PDF.

\subsection{IllustrisTNG}
The IllustrisTNG simulation suite \citep{2019ComAC...6....2N,2018MNRAS.475..648P,springel_first_2018,2018MNRAS.480.5113M} is a state-of-the-art cosmological hydrodynamical simulation that employs a moving-mesh technique based on Voronoi cells, with resolution that adapts to local density. Thanks to its subgrid physics models, the simulation incorporates key physical processes such as star formation, radiative metal cooling, stellar and supermassive black hole feedback, and chemical enrichment from Type II and Type Ia supernovae and AGB stars \citep[for more details, see][]{2018MNRAS.473.4077P}.
To measure the DM, the electron density along the LOS is required. Previous studies \citep{Jaroszy_DM_2020, 2021ApJ...906...49Z, takahashi_statistical_2021} have already used simulations to create DM maps.
In particular, \citet{2024A&A...683A..71W,2021ApJ...906...49Z,Mo_2022} studied the DM PDF as obtained from IllustrisTNG. However, \citet{2025arXiv250707090K}\footnote{\href{https://ralfkonietzka.github.io/fast-radio-bursts/ray-tracing-catalogs/}{https://ralfkonietzka.github.io/fast-radio-bursts/ray-tracing-catalogs/}} showed that some of those results suffer from periodic boundary conditions and an insufficient treatment of Voronoi cells. \review{We will therefore use the results from \citet{2025arXiv250707090K}, measured at $z=0.7, 1.5, 3, 4, 5$, as our comparison baseline. Their analysis is based on the TNG300 realization of the IllustrisTNG suite, in particular TNG300-1, which provides a large simulation volume suitable for capturing the impact of large-scale structure on FRB DMs.}
\vspace{.3cm}

\section{Results}
\label{sec:Results}
In this section, we describe the results of the comparison in detail. To this end, we first compare with the consistency-check simulations in Section \ref{sec:comp_cons_sim} to provide a consistency check and an estimate for the covariance. We then compare the results to IllustrisTNG in Section \ref{sec:comp_TNG} and demonstrate how the PDF depends on the BFC parameters in Section \ref{sec:bcmpara}. Lastly, we use our results to investigate the assumption of log-normality for the FRB DM PDF in Section \ref{sec:snr}.

\subsection{Comparison with the consistency-check simulation}
\label{sec:comp_cons_sim}

In \Cref{fig:simulation_comp}, we present a comparison between the analytical calculation given in \Cref{eq:PDF_without_clustering} and the simulation results. Note that this result is obtained without clustering, since the consistency simulation randomly places halos.
Both methods use the same BFC parameters, with the FRB redshift fixed at $z = 1.5$. The mass range considered spans $10^{10} \; h^{-1} M_\odot$ to $10^{13} \; h^{-1} M_\odot$. 
The shaded region in \Cref{fig:simulation_comp} represents the $3\sigma$ confidence interval computed from the sample covariance. We find good agreement between the simulation and the analytical prediction. The remaining discrepancies can be attributed to two closely related effects. First, both the LOS and halo masses are divided into a finite number of bins, with fixed redshift and mass assigned to all objects within each bin. Increasing the number of bins would improve accuracy and more closely align the simulation with the analytical result.

Second, the dominant contribution to the DM arises from haloes with masses $> 10^{12} \, h^{-1} M_\odot$. As the number of bins increases, the probability of sampling such massive haloes within a single bin decreases. To ensure adequate statistical sampling of these rare systems, larger bin volumes are required. However, increasing the bin size, which corresponds to larger impact parameters $R_\mathrm{Im}$, significantly raises the computational cost of the simulation, especially when running many realisations for the covariance estimation.  

We have verified that those two effects indeed explain the remaining discrepancy between the simulation and our analytical results.

\begin{figure}
    \centering
    \includegraphics[width=\linewidth]{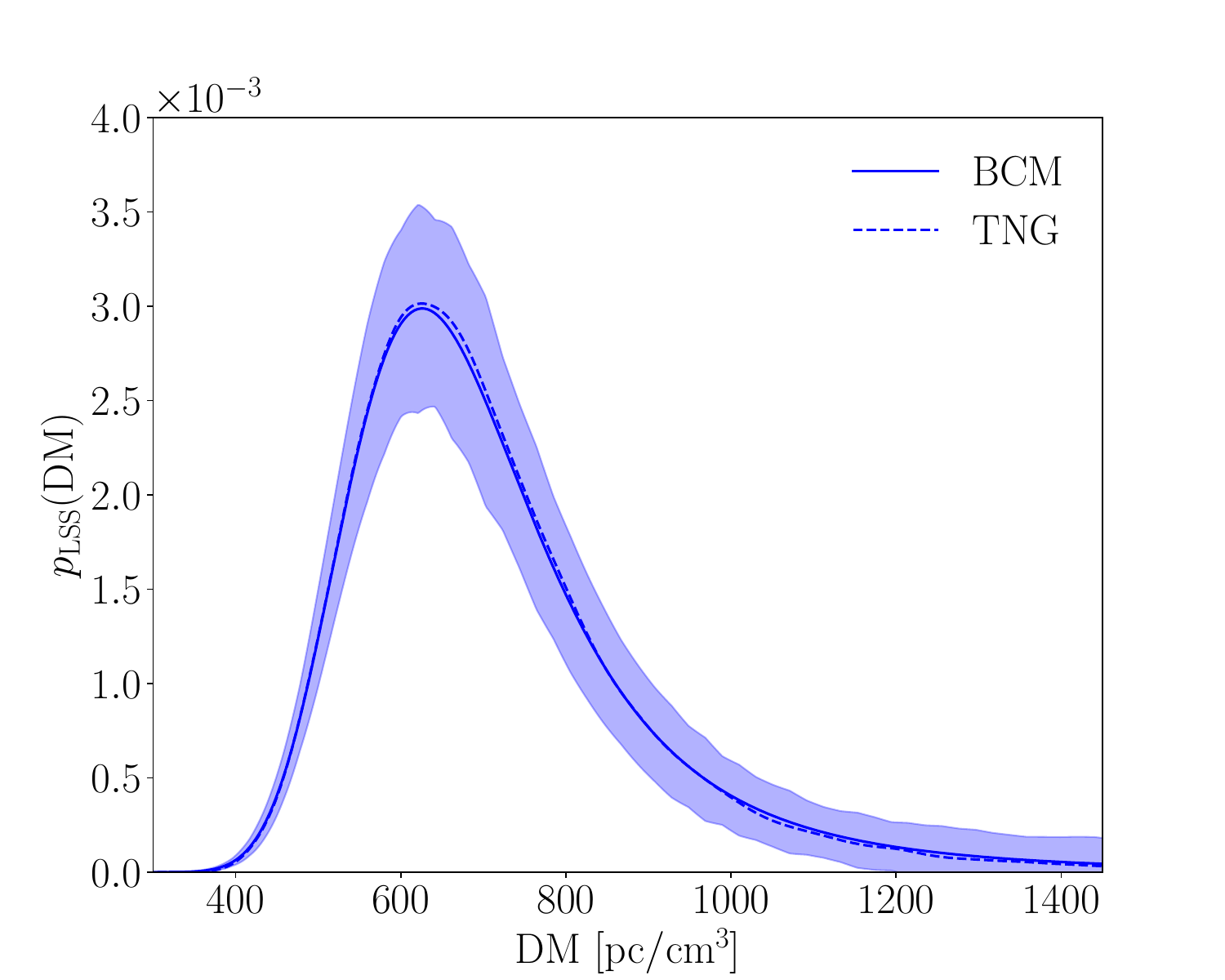}
    \caption{\review{Comparison between the TNG PDF at redshift 0.7 and the PDF obtained from the BFC model using the best-fit parameters derived from the profiles at the same redshift. The solid line shows the BFC PDF, and the dashed line shows the TNG PDF. The shaded region represents the $3\sigma$
confidence interval calculated using the covariance matrix. Crucially, the solid line is a prediction based on the hot gas profiles (see \Cref{fig:fit_to_profile}) rather than a fit to the dashed line.}}
    \label{fig:pdf_profile}
\end{figure}

\subsection{Comparison with IllustrisTNG}
\label{sec:comp_TNG}

\subsubsection{From profiles to the PDF}
\review{The first question we would like to answer is whether the model presented here is self-consistent in the following way: Suppose we fit the gas profiles in IllustrisTNG with the BFC model, as done in \citet{2025arXiv250707892S}, can we obtain a consistent prediction for the PDF of the DM? To this end, we use that fit to the gas profiles at redshift 0.5 and show the results in \Cref{fig:fit_to_profile}. As expected, the best-fit BFC profiles reasonably well describe the IllustrisTNG measurements over a broad mass range. We can now use the derived best-fit parameter values and compute the associated DM PDF. The result is shown in \Cref{fig:pdf_profile}, and as can be seen, the BFC prediction matches the simulation measurement very well, particularly relative to the cosmic-variance uncertainty (shaded region). It is important to highlight that the solid line in \Cref{fig:pdf_profile} is not a fit to the dashed line but a prediction. 
This is a nontrivial result, indicating that BFC provides a self-consistent description of halo gas profiles and integrated quantities, such as the DM PDF. }The result is noteworthy for two reasons: $(i)$ the DM is an integrated effect while the profiles are only fitted at a particular redshift, $(ii)$ the DM accumulates contributions from all halo masses, with the weight at a given halo mass given by the number density. Furthermore, previous studies \citep[see e.g.][]{2025NatAs.tmp..131C,2025ApJ...983...46M,2024A&A...683A..71W} have found that the halo-model-based approach to FRBs is challenging, as DM is highly sensitive to gas in filaments rather than in halos. Our results demonstrate that this description is indeed self-consistent, at least at the level of the PDF. This is further supported by the fact that halo clustering itself is not particularly important. For other observables, e.g. the power spectrum, the role of filaments might be more important, especially in the transition regime between the one- and two-halo term. This could be alleviated by using a web-based halo model as in \citet{2025arXiv250810902B}, or by introducing a geometric mean between the one- and two-halo term, as for example done in \citet{mead_accurate_2015}.

\begin{figure}
\centering
\includegraphics[trim=1.5cm 0 2.5cm .2cm, clip,width=.5\textwidth]{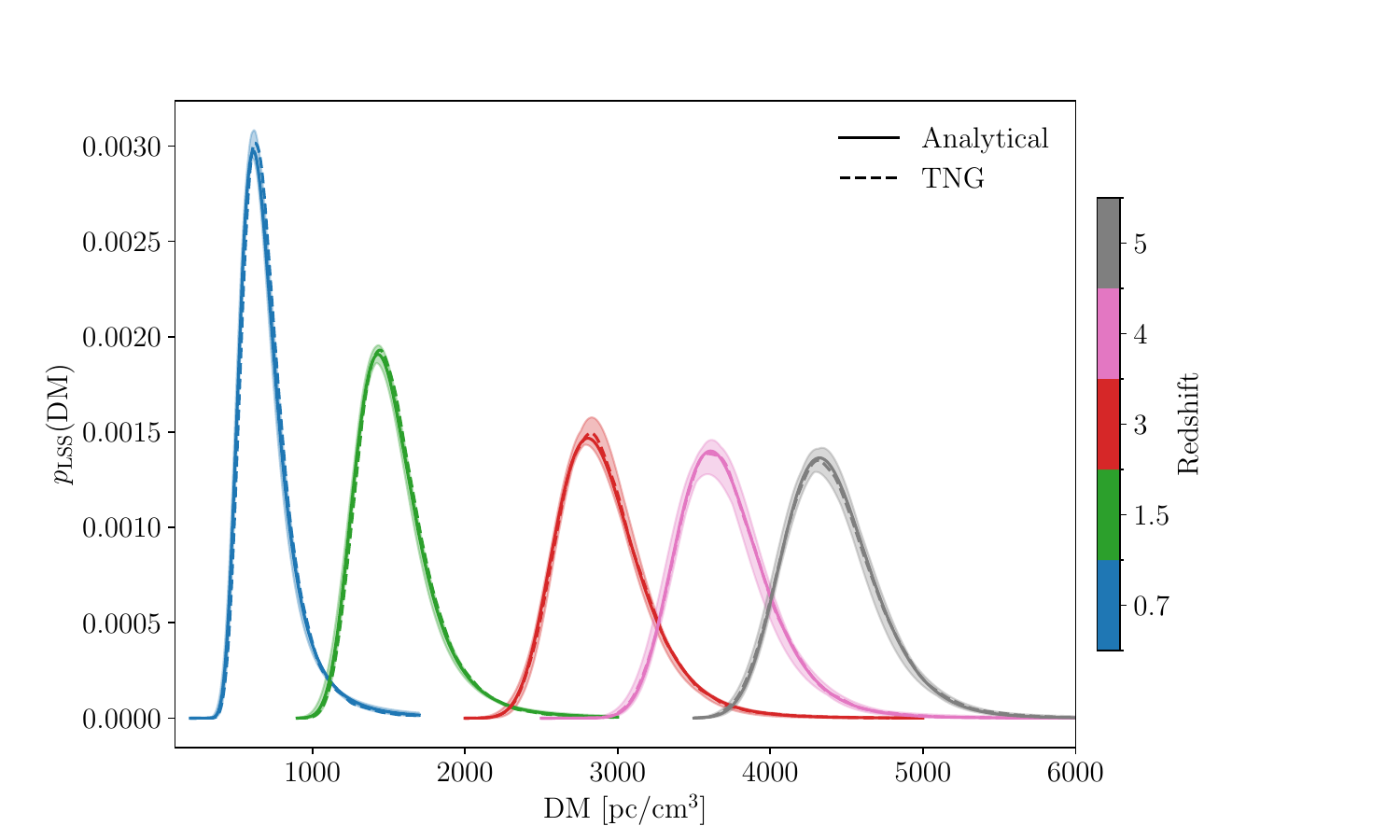}
\caption{Comparison of the analytical PDF for the DM of FRBs with results from the TNG simulation across various redshifts. The solid lines represent the analytical predictions, while the shaded regions indicate the $68 \%$ confidence intervals. The dashed lines show the corresponding results from the TNG simulation.}
\vspace{.05cm}
\label{fig:result1}
\end{figure}

\begin{figure*}
\centering
\includegraphics[width=\linewidth]{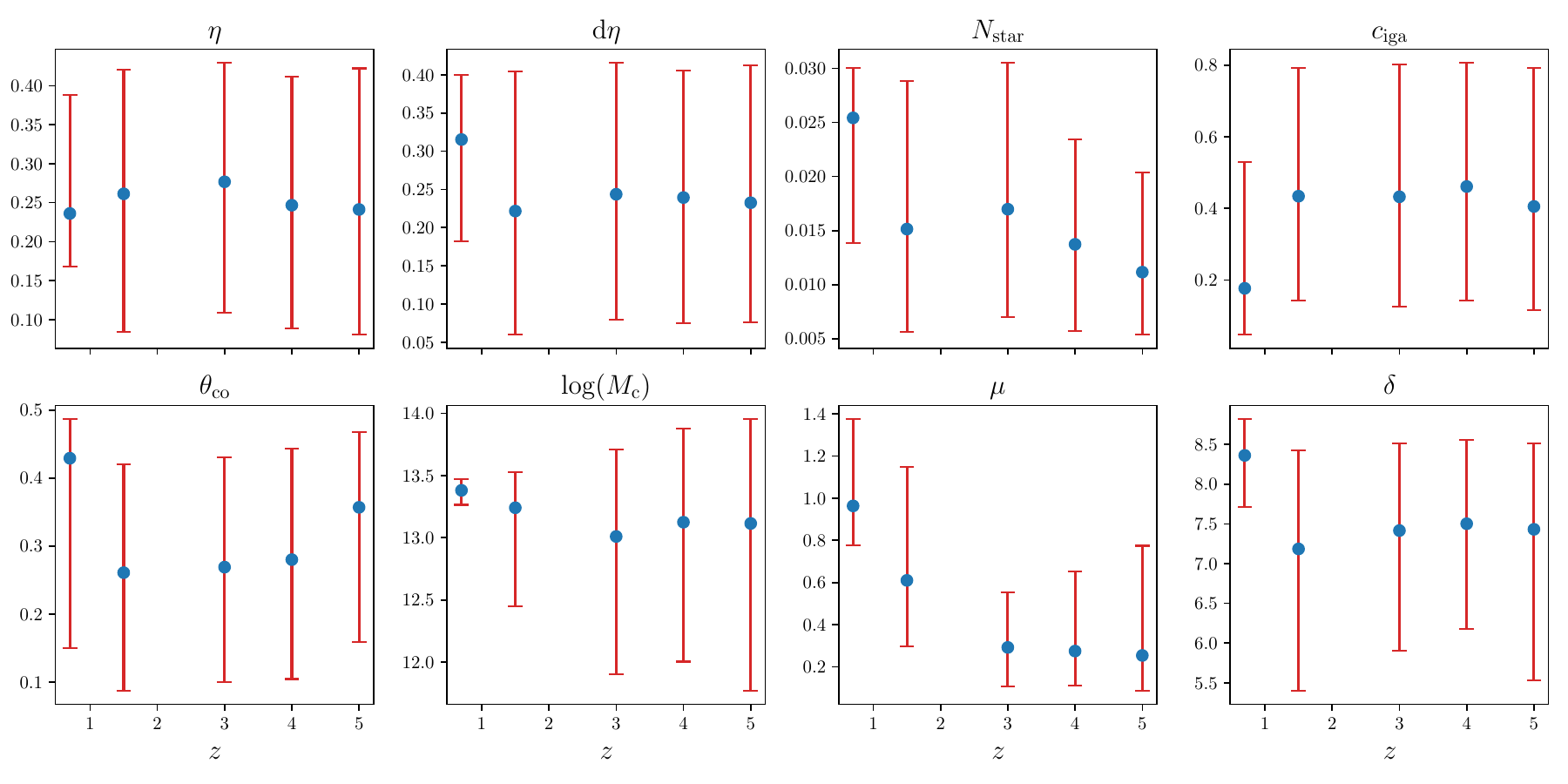}
\caption{Model parameters obtained from fitting to the IllustrisTNG simulation as a function of redshift for $\sim 10^4$ FRBs. Red error bars indicate the 68\% confidence intervals, and blue points mark the highest posterior probabilities.}
\label{fig:MCMC}
\end{figure*}

\begin{figure}
    \centering
    \includegraphics[width=\linewidth]{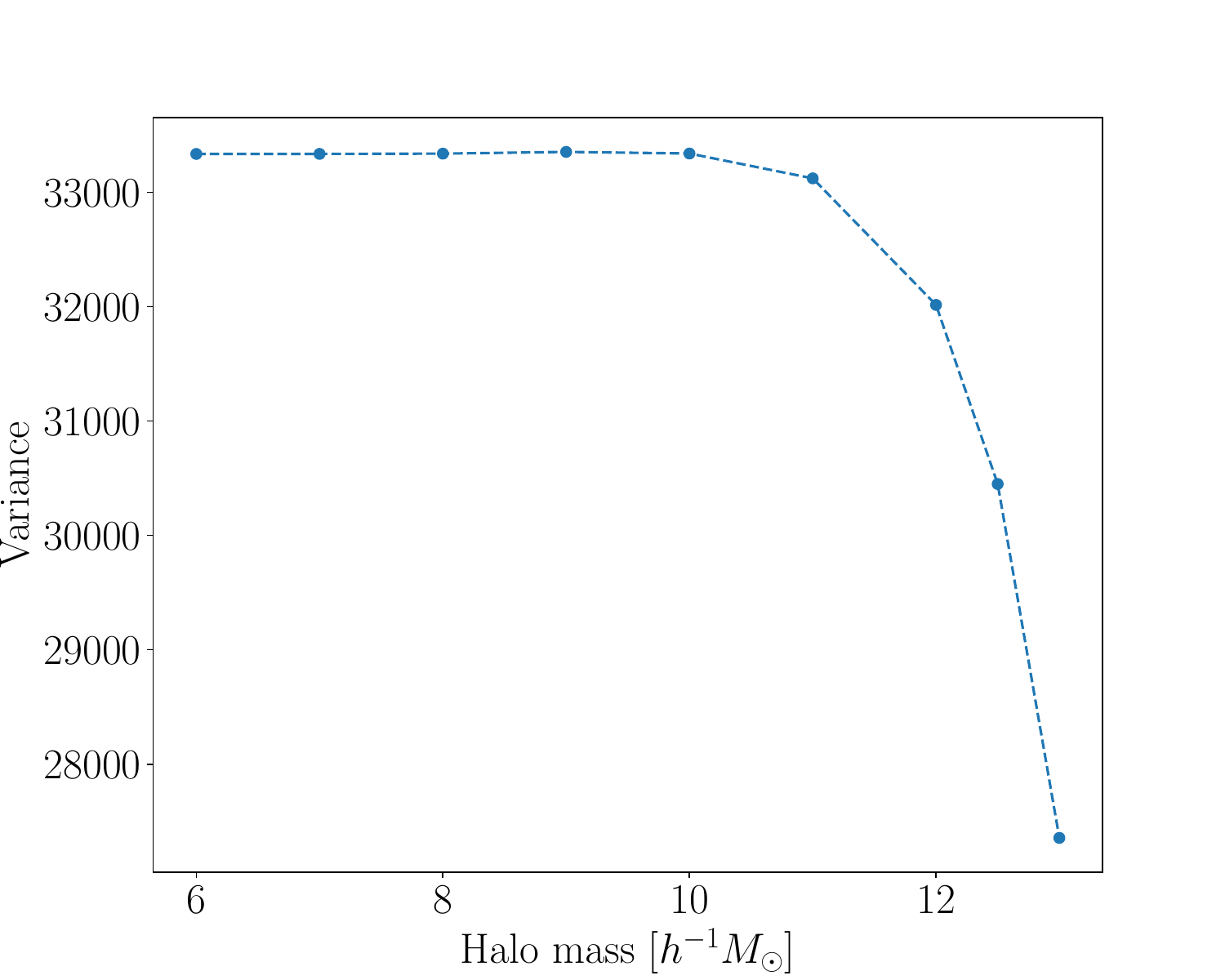}
    \caption{Dependency of the PDF variance on the lower mass bound at a fixed redshift 0.7.}
    \label{fig:resultlowermassbound}
\end{figure}

We investigate this last point in more detail and show how the PDF variance depends on the lower mass bound for all integrals over halo mass in \Cref{fig:resultlowermassbound}. As the lower mass bound decreases, the variance increases, indicating a broader PDF, which is expected. Including low-mass halos in the calculation naturally increases the DM contribution. However, the contribution of an individual low-mass halo is minimal, so we do not expect a drastic impact on the PDF. On the other hand, the halo mass function predicts a large number of low-mass halos, which compensates for their small individual contributions. 
It is worth noting that, beyond a certain point, further lowering the mass bound does not change the PDF variance. This occurs because halos in that mass regime are too small and have shallow gravitational potentials, allowing feedback processes to expel their gas essentially entirely. With no remaining gas, these halos no longer contribute to the DM. This is, of course, a simplified picture, as very low-mass halos will not host active galactic nuclei and therefore have less feedback. All the relations used in our model are calibrated to halos with masses $M\gtrsim 10^{11}\,h^{-1}M_\odot$. One might therefore raise the suspicion that the plateau at low masses seen in \Cref{fig:resultlowermassbound} is somewhat an artefact of the BFC rather than a genuine physical feature. However, we tested what happens if halos retain an unrealistic amount of gas at low masses, in particular, a mass-independent gas fraction $f_\mathrm{hga} = f_\mathrm{b}$. Even in this case, variance saturates at $10^{8}\, h^{-1} M_\odot$ and only changes by around $5\;\%$ relative to $10^{10}\, h^{-1} M_\odot$. 
Therefore, even if halos at lower masses would retain their gas, the result would not change much, as the higher abundance of low-mass halos can not compensate for the drop in the gas density at some point. This finding is in line with Figure 1 in \citet{2025ApJ...989...81S}. 
Throughout this paper, we therefore restrict our analysis to halos in the mass range \review{$10^{8}\, h^{-1} M_\odot$ to $10^{16} \,h^{-1} M_\odot$}.

\subsubsection{Fitting the PDF directly}
\review{Now that we have established that the model is self-consistent, i.e., that the PDF indeed follows from the profiles without further fitting, we are interested in how well the BFC parameters can be constrained.
To compare the model to the TNG simulation, we first define a likelihood using a covariance matrix. We bin both the TNG PDF and the BFC PDF into $N_\mathrm{bin} = 100$ bins and use the covariance matrix from our consistency-check simulation, evaluated at the fiducial parameter values in \Cref{tab:priorMCMC} for the BFC. We assume a Gaussian log-likelihood}
\begin{equation}
\label{eq:loglike}
    \log(\mathcal{L}) = -\dfrac{1}{2}\sum_{i,j =0}^{N_\mathrm{bin}} \left( p^\mathrm{TNG}_i - p^\mathrm{BFC}_i \right) \textbf{C}^{-1}_{ij} \left( p^\mathrm{TNG}_j - p^\mathrm{BFC}_j\right), 
\end{equation}
where $p^\mathrm{TNG}_i$ and $p^\mathrm{BFC}_i$ are the PDF values in bin $i^\mathrm{th}$ for the TNG and BFC, respectively. Note that the covariance we are using here effectively assumes infinitely many FRBs.

We then sample the likelihood in a Markov Chain Monte Carlo (MCMC) analysis using the ensemble sampler {\sc emcee} \citep{2013PASP..125..306F}. We choose to fit each of the five TNG PDFs at its own redshift. This is due to the fact that the BFC model has mostly been tested and calibrated at redshifts lower than $z=1.5$, and some model parameters show significant redshift dependence in this regime \citep[see e.g.][]{2025arXiv250707892S}. Hence, fitting all redshifts simultaneously could yield erroneous results due to potential redshift-dependent parameter evolution. The prior ranges used for all redshift bins are identical and are listed in \Cref{tab:priorMCMC}.

The best-fit PDFs derived from the MCMC are shown in \Cref{fig:result1} along with their $68\,\%$ confidence intervals. The results are colour-coded by redshift; the TNG measurements from \citet{2025arXiv250707090K} are shown in dashed, while the best-fit analytical results from BFC are shown as a solid line. 
As shown, the analytical model agrees well with the hydrodynamical simulation. \review{To quantify this agreement, we perform a $\chi^2$ test for each fit relative to the cosmic variance using the covariance matrix. The resulting values for the reduced $\chi^2$ are 1.51, 0.30, 0.22, 0.40 and 1.17 for the fits at redshifts 0.7, 1.5, 3, 4, and 5, respectively. It is important to keep in mind, however, that the data we are fitting the BFC to is already a fit itself against the results of the TNG simulation. This means that the covariance we are using might not be entirely representative of the scatter in that data. Therefore, interpreting the reduced $\chi^2$ has to be done with caution.
} 
The corresponding constraints on the parameters of the BFC model are demonstrated in \Cref{fig:MCMC}. 
We note again that we performed the fit at each redshift separately, thus allowing us to investigate redshift evolution of the parameters. As can be seen, the primary drivers of the PDF shape are the parameters determining the gas profile, \Cref{eq:rhogas}. These are $M_\mathrm{c}$, $\mu$ and $\delta$.
Interestingly, we find that $\mu$ depends strongly on redshift, favouring a mass-independent slope of the gas profile, $\beta$, at higher redshifts, since $\mu\to 0$ implies a constant $\beta$ (compare \Cref{eq:beta}). A large $\mu$ will introduce a stronger dependence on $M_\mathrm{c}$. Compared to the parameters obtained in \citet{2025arXiv250707892S} for redshifts 0.7 and 1.5, we find that $\delta$ is larger and $\mu$ is less dependent on redshift. However, one should note that the redshift range here is much larger.
Finally, we note that $\eta$ and d$\eta$ are not constrained when compared to the prior range in \Cref{tab:priorMCMC}. These parameters describe the stellar and central galaxy profiles, which are scaled by $N_\mathrm{star}$. As the best-fit values obtained for $N_\mathrm{star}$ are small, the exact shape of the profile is irrelevant, and thus the influence of $\eta$ or d$\eta$ is negligible. The same is true for the cold gas component, $c_\mathrm{iga}$. Comparing the resulting error bars to the constraints of the BFC model obtained by kSZ and X-ray observations \citep{2025arXiv250707991K,2025ApJ...989...81S} shows that $\sim10^4$ FRBs can constrain the BFC parameters to similar precision.

\begin{figure}
    \centering
    \includegraphics[trim=1.1cm 0 3.5cm .2cm, clip,width=.5\textwidth]{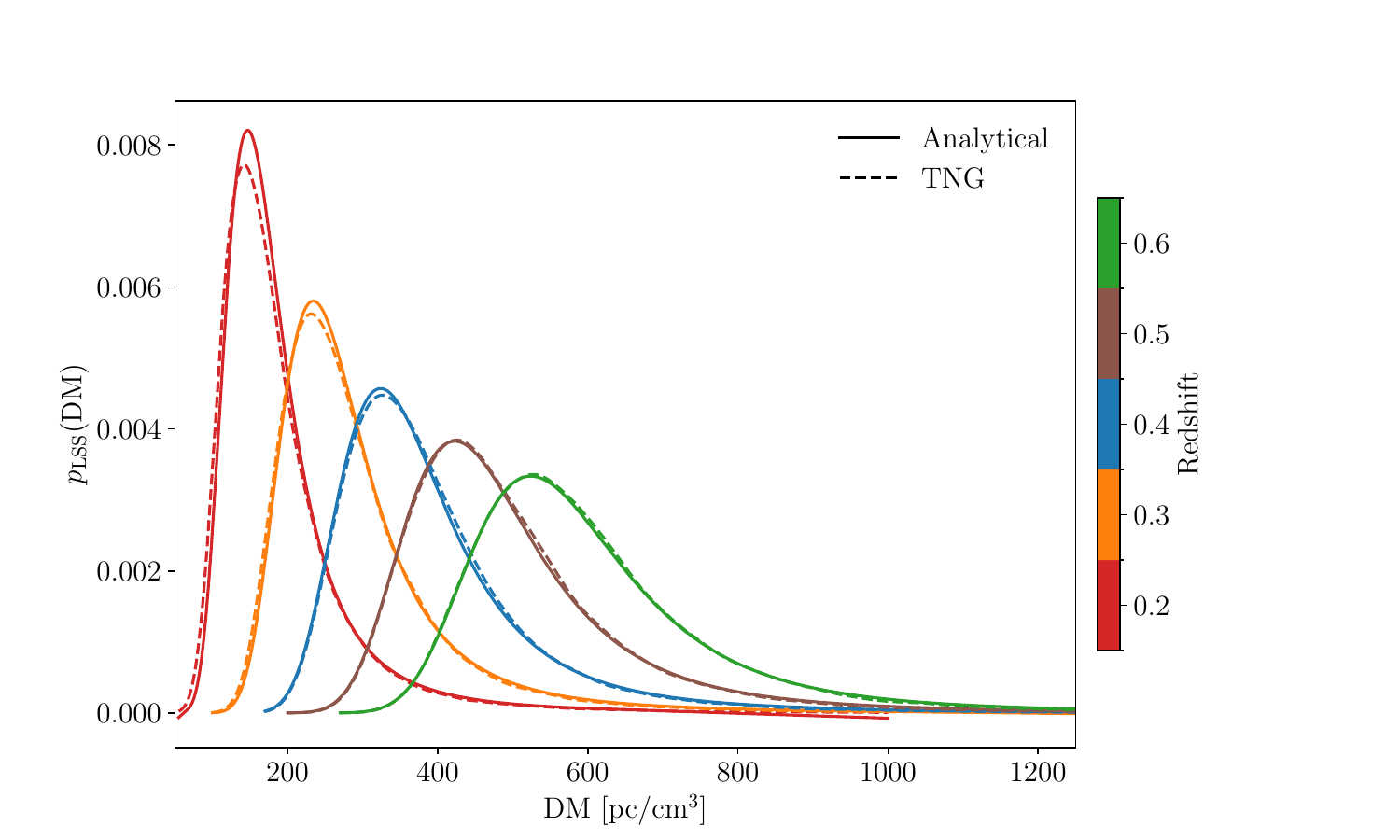}
    \caption{Comparison of the analytical PDF for the DM of FRBs with results from the TNG simulation across redshifts from $0.2$ to $0.6$. Solid lines show the analytical results with fiducial parameters, while dashed lines represent the results from the TNG simulation.}
    \label{fig:lowredshift_pred}
\end{figure}

\begin{figure*}
    \centering
    \includegraphics[width=\linewidth]{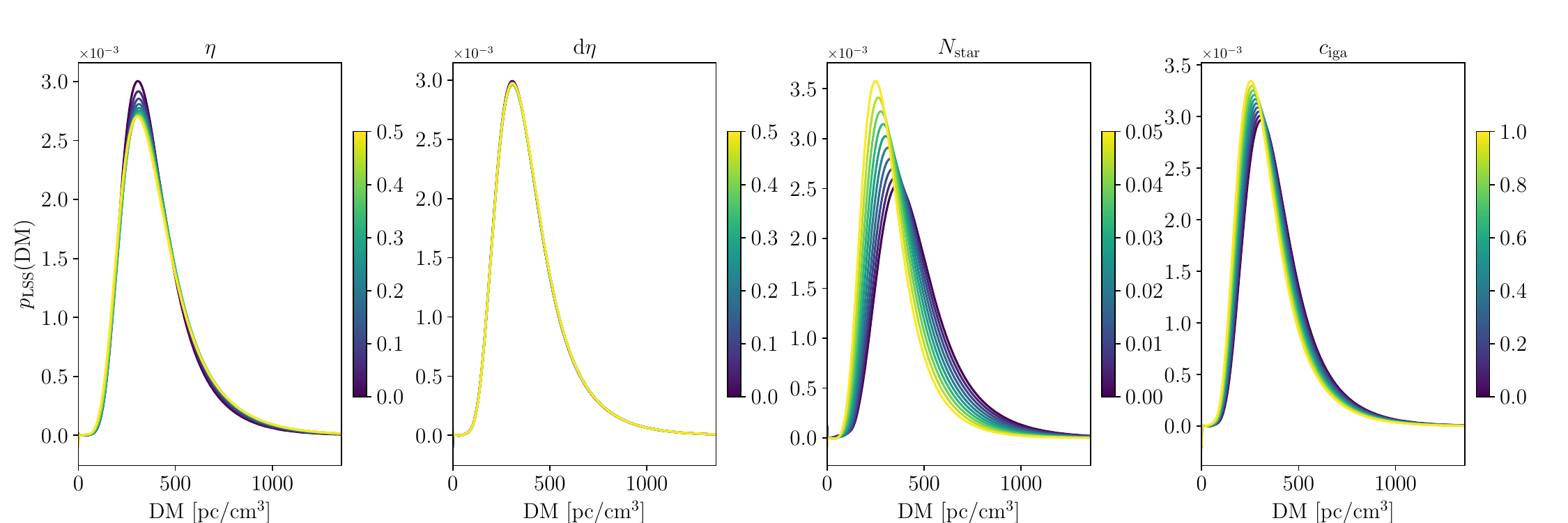}
    \includegraphics[width=\linewidth]{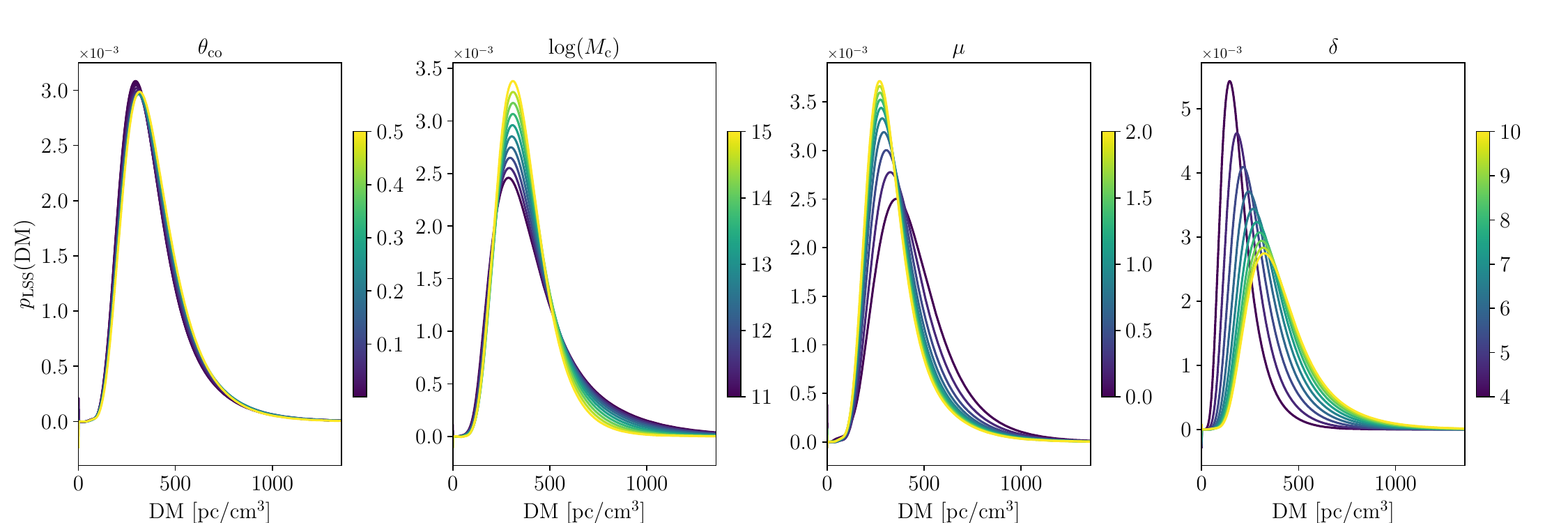}
    \caption{Impact of varying individual BFC parameters on the predicted DM PDF at redshift 0.7. In each panel, only one parameter is varied at a time, with values indicated in the colourbar. All the others are held at their fiducial values shown in \Cref{tab:priorMCMC}. Parameters with a direct influence on the gas distribution, such as $M_\mathrm{c}$, $\mu$, and $\delta$, produce more significant changes in the shape of the PDF.}
    \label{fig:BCMdep}
\end{figure*}

\subsubsection{PDF at low redshifts}

In this section, we compare our results with the TNG simulation at low redshifts ($z = 0.2\text{--}0.6$). We adopt the fiducial parameters listed in \Cref{tab:priorMCMC} and compute the corresponding PDFs at each redshift. The results are presented in \Cref{fig:lowredshift_pred}.

The BFC model provides a good match to the simulation down to $z \approx 0.3$, but begins to deviate at lower redshifts. Two main factors likely contribute to this behaviour. First, as indicated in both previous BFC studies \citep[see e.g.][]{2025arXiv250707892S} and our result (see e.g. \Cref{fig:MCMC}), the BFC parameters are not strictly redshift-independent. Using a fixed parameter set across different redshifts can therefore introduce systematic discrepancies. Second, as redshift decreases, the host contribution becomes increasingly significant. While our model accounts only for the LSS contribution to the PDF, the TNG simulation includes both LSS and host contributions, which can lead to additional differences at low redshift.

\subsection{Effect of the BFC Parameters}
\label{sec:bcmpara}
One of the main benefits of BFC is that one can study how different observables (here, the DM PDF) depend on physically motivated feedback parameters and, hence, on the underlying density profiles. This is almost impossible in hydrodynamical simulations. If one, for example, can identify parameters which are irrelevant for cosmological studies, the complexity of the BFC model can be reduced. In this section, we therefore examine how the DM PDF depends on various BFC parameters. 

To isolate the effect of each parameter, we fix all others to their fiducial values (i.e., the best-fit parameter for TNG as shown in \Cref{tab:priorMCMC}) and vary only one at a time. Note that we do not fix the mean of the corresponding distributions.
The results are shown in \Cref{fig:BCMdep}, and the observed trends are consistent with the MCMC results presented in \Cref{fig:MCMC}. 

In BCM, there are three parameters, $M_\mathrm{c}$, $\mu$, and $\delta$, that directly influence the gas profile, and as can be seen from \Cref{fig:BCMdep}, these have the largest effect on the DM PDF. Increasing parameters such as $M_\mathrm{c}$ and $\mu$ makes the gas profile shallower, indicating that less gas remains concentrated within the halo, i.e. stronger feedback. This corresponds to a more homogeneous distribution of gas, which reduces the variation in DM along different lines of sight at a fixed redshift, making the PDF more sharply peaked. This can be compared to the limiting case of a perfectly homogeneous gas distribution, which results in a Dirac delta PDF equal to the mean.
The parameter $\delta$, however, has the opposite effect: increasing it 
 makes the gas more concentrated inside the halo, resulting in a broader PDF.

On the other hand, parameters such as $\eta$ and $\mathrm{d}\eta$, which influence the stellar contribution, have only a small effect on the PDF. The reason is that they solely influence the stellar mass fraction as a function of halo mass, which in turn is scaled by $N_\mathrm{star}$. So these parameters induce a secondary effect on the stellar profile, which only indirectly changes the gas profile via \Cref{eq:rhogas}. Generally, if $\eta + \mathrm{d}\eta$ is larger, the stellar fraction decreases more rapidly with halo mass, leading to a larger gas fraction and, consequently, a slightly broader PDF. This can be seen as follows: if the gas fraction is multiplied by some factor $f$, the random variable DM is simply rescaled $\mathrm{DM}\to f\mathrm{DM}$, as are the corresponding moments. Thus, if this factor is larger than unity, the PDF becomes broader.
For $N_\mathrm{star}$, this influence is more direct: an increase leads to less gas in the halos because it is locked up in stars and hence to a narrower PDF, again by rescaling the random variable DM. Adding more cold gas by increasing $c_\mathrm{iga}$ has the same effect.
We also see relatively little influence of $\theta_\mathrm{co}$ as only those sightlines that pass through the halo centre are sensitive to this parameter. 
However, if feedback is weak overall, the gas is more centrally concentrated, and therefore significant contributions to the DM occur only close to the halo centre. Hence, one can expect the PDF to be more sensitive to $\theta_\mathrm{co}$ if feedback is weak.

\subsection{The case of the log-normal distribution}
\label{sec:snr}
In this section, we test whether the DM PDF measured from IllustrisTNG can be well-described by a log-normal distribution. \citet{2025arXiv250707090K} indicated that, at low redshift, the PDF is not best described by a log-normal distribution.
For this purpose, we take the results of the TNG simulation at redshift  $z=0.7$ as our reference dataset and fit it with a log-normal \citep[compare to Figure 13 in][]{2025arXiv250707090K}. Alternatively, we could fit the log-normal variance of the corresponding PDF directly to the BFC parameters. However, we are mainly interested in the goodness-of-fit for the log-normal as it is commonly used in FRB analyses.
The question we hence want to answer is: how many FRBs are required to distinguish the log-normal fit from the real (IllustrisTNG) distribution?

To this end, we perform a $\chi^2$ test. We again assume a Gaussian likelihood, \Cref{eq:loglike}, where $p^\mathrm{BFC}_i$ is now replaced by $p^\mathrm{LN}_i$.
To compute the covariance matrix, following the procedure in the previous section, we first divide the PDF into 100 bins and then run multiple realisations of our consistency simulations to estimate the cosmic-variance contribution as before.
To account for the finite number of FRBs, we add noise to each bin by drawing a random number from a normal distribution with zero mean and variance:
\begin{equation}
    \sigma^2_\mathrm{noise} = \frac{p_i}{{N_\mathrm{FRB}\Delta\mathrm{DM}_i}},
\end{equation}
 where $p_i$ is the PDF value in the 
$i$-th bin and $\Delta\mathrm{DM}_i$ is the corresponding DM width. This approach effectively models the increased uncertainty at low FRB counts: for small $N_\mathrm{{FRB}}$, the fluctuations are large, while at the limit
$N_\mathrm{FRB} \rightarrow \infty$ this contribution to the uncertainty vanishes. In principle, this effect is included in our simulation. However, we choose to estimate the covariance from many sightlines $N_\mathrm{FRB}\to\infty$ and add this contribution by hand. From the Gaussian statistic, we can immediately estimate the $\chi^2$ and hence the $p$-value given the degrees of freedom.

 The previous approach is only valid under the assumption of a Gaussian likelihood, which does not hold in our case. For consistency, we also adopt the following procedure. 
We begin by drawing $N_{\mathrm{FRB}}$ samples from the TNG simulation. We then approximate these sampled values with a log-normal distribution by maximising the likelihood of the data. In practice, this is done by estimating the parameters of the log-normal PDF, \(p^{\mathrm{LN}}\), that maximise
\begin{equation}
    \label{eq:loglike_lognormal}
     \log(\mathcal{L}) = \sum_{i=1}^{N_\mathrm{FRB}} \log\left[p^\mathrm{LN}\left(\mathrm{DM}_i\right)\right]
\end{equation}
where $\mathrm{DM}_i$ are the dispersion measures drawn from the TNG simulation.

\begin{figure}
    \centering
\includegraphics[width=0.49\textwidth]{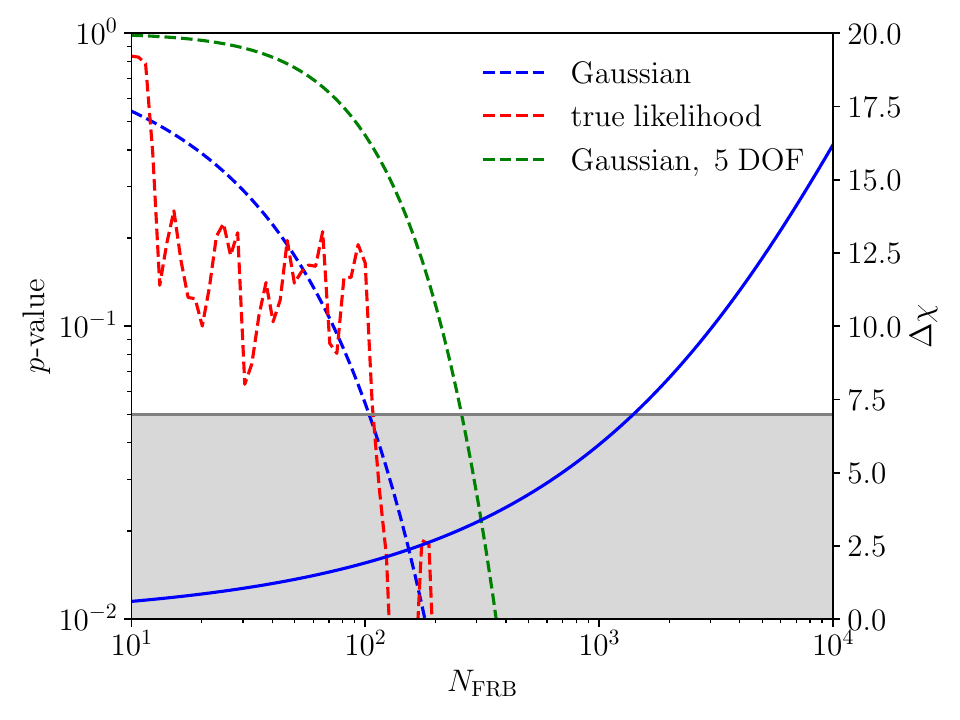}
    \caption{Goodness-of-fit as a function of FRBs observed, $N_\mathrm{FRB}$. The dashed curves show the $p$-values. Blue and green show the $p$-value obtained from the Gaussian likelihood assumption, \Cref{eq:loglike}, for one and five degrees of freedom (DOF), respectively. In red we show $1-p_\mathrm{te}$, \Cref{eq:pte}, thus dropping the Gaussian likelihood assumption.
    The solid curve indicates the corresponding $\Delta\chi$ for the Gaussian case.
    We show the typical $p$-value threshold of $p=5\times 10^{-2}$ as the grey shaded area.}
    \label{fig:ch2logn}
    \vspace{.3cm}
\end{figure}

 We estimate the $p$-value as follows: we generated $N$ random realisations from the likelihood evaluated at the maximum posterior. These realisations serve as an alternative to the usual $\chi^2$ statistics. By comparing the log-likelihood of the actual data, $\chi_{\mathrm{data}}^2$, with those of the simulated realisations, we can estimate the one-sided $p$-value (or probability-to-exceed) as
\begin{equation}
\label{eq:pte}
p_\mathrm{te} = P \left( \chi_i^2 > \chi_{\mathrm{data}}^2 \right).
\end{equation}
\Cref{fig:ch2logn} shows how the $\Delta\chi^2$ (solid blue curve) for the Gaussian likelihood and the corresponding $p$-values (blue dashed) as a function of $N_\mathrm{FRB}$. For the latter, we assume that we fit only a single parameter, i.e., the amplitude of the DM-redshift relation. The corresponding version of the exact likelihood, i.e. \Cref{eq:loglike_lognormal,eq:pte}, is shown as the red dashed line. The noise occurs because of the random draws of the data. One can, however, see that the overall behaviour is the same and that the $p$-value drops below the commonly accepted value of $5\times 10^{-2}$ at around $10^2$ FRBs (as indicated by the grey-shaded area). If we assume that a typical analysis uses at least five free parameters\footnote{This could be a couple of cosmological parameters and two parameters for the host contribution.}, we can see from \Cref{fig:ch2logn} that the $p$-value drops below $5\times 10^{-2}$ for a few hundred FRBs. Hence, using the log-normal approximation for the PDF is sufficient unless more than a few hundred FRBs are being analysed. Coincidentally, this is also the number of FRBs at which correlations between the different sightlines will bias parameter estimation as well \citep{reischke_cosmological_2023}.

\subsection{Possible limitations}
\review{
Despite the encouraging results of this study, our approach has some potential limitations that we discuss here. In general, the two-halo term not only determines the mean of the DM PDF but also affects its overall shape. Far outside the halo, the electron density field approaches the mean electron density of the Universe and therefore primarily shifts the mean DM density. However, the transition region between the one-halo and two-halo terms, an inherent limitation of halo-based models, can affect the shape of the PDF.
In this study, we considered only the one-halo term (though clustering of those halos is included). The very good agreement with the TNG simulation results suggests that the two-halo term in the shape of the PDFs is not dominant.}

\review{
Another limitation of our work is the redshift dependence of the BFC parameters. Ideally, we would like to identify a single set of BFC parameters that can describe the baryonic distribution of the Universe over a broad redshift range. However, both \citet{2025arXiv250707892S} and our results indicate that some of these parameters vary with redshift.
This can potentially be problematic because the DM PDF is an integrated quantity that depends on electron density along different sightlines out to a given redshift. Any redshift dependence of the BFC parameters, therefore, introduces additional uncertainty and affects the reliability of our results.}

\review{Lastly, our work relies on a single hydrodynamic simulation and hence a single feedback prescription. In the case of a simulation with stronger feedback, where gas profiles are even shallower, the two-halo term, as well as the integration limit (see Appendix \ref{app:rmax}), could potentially change as a function of feedback. However, we reiterate that the choices made here, such as the maximum integration radius and the inclusion of the two-halo term, were made for numerical reasons rather than to improve agreement between the simulation and the analytical approach. Together with the fact that the analytical model matches the DM PDF very well without directly fitting it (see \Cref{fig:pdf_profile}), this is achieved by adjusting the profiles, giving us additional confidence in the generality of our approach. A more detailed investigation of the limitations discussed above will be carried out in future work.}

\vspace{.3cm}
\section{Summary \& Conclusion}
\label{sec:Conclusion}
In this work, we present a fast, analytical framework for predicting the one-point PDF of the DM for FRBs using a baryonification (BFC) model. The theoretical modelling was tested and validated against consistency simulations and full hydrodynamical simulations using IllustrisTNG. Our model provides an efficient alternative to expensive hydrodynamical simulations for studying FRBs in a cosmological setting. 
To accelerate subsequent calculations, we used {\sc{cosmopower}} to train an emulator that enables rapid evaluations across parameter space without recalculating the full analytical expression for each parameter set. The principal results of the paper can be summarised as follows:

\begin{enumerate}
    
    \item[i)] \review{We demonstrated the self-consistency of our approach by fitting gas density profiles from the TNG simulation with the BFC model and subsequently using the best-fit parameters to predict the DM PDF. This confirms that the BFC model produces physically consistent descriptions of both local gas distributions and integrated statistics such as the DM PDF.}

    \item[ii)] \review{The model is flexible enough to reproduce the IllustrisTNG simulation at five redshifts ($z$ = 0.7, 1.5, 3, 4, and 5) and shows excellent agreement between the analytical predictions and the hydrodynamic simulation results. We see, however, that some of the BFC parameters exhibit a redshift dependence when fitted at each redshift individually, suggesting that the BFC parametrisation may require an explicit redshift dependence, in particular for $\mu$ or $M_\mathrm{c}$. These findings are consistent with those reported by \citet{2025arXiv250707892S}.}
    
    \item[iii)] We identified the BFC parameters that determine the PDF shape: the pivot mass scale $ M_\mathrm {c} $, the mass-dependent slope parameter $\mu$, and the outer transition parameter $\delta$. These parameters, which directly control the gas profile, emerged as the primary drivers of the PDF morphology, whereas parameters describing the stellar and cold gas components had only minor effects. A survey with $\sim 10^4$ FRBs can constrain the gas parameters of the BFC model to similar precision as 
    kSZ and X-ray observations \citep[see as well][]{2025arXiv250707991K,2025ApJ...989...81S}. Therefore, FRBs provide an independent measurement of the gas distribution, sensitive to halo mass and to systematics.
    
    \item[iv)] We used our pipeline to determine how well the FRB DM PDF can be described by a log-normal distribution, and for which number of FRBs this approximation breaks down \citep{2025arXiv250707090K}. We demonstrated that while this simple functional form provides an adequate description for a few hundred FRBs, the approximation becomes distinguishable from the true PDF for larger samples. 

\end{enumerate}
\review{
We would like to reiterate that gas density profiles fitted to hydrodynamic simulations can accurately predict the PDF of the DM without fitting the PDF itself, thereby providing a non-trivial validation of the halo model approach for FRBs.} This result suggests that FRBs, despite their sensitivity to gas distributed throughout the circumgalactic and intergalactic media rather than just within virialised halos, can still be modelled within a halo-centric framework when considering integrated quantities. This opens the door for consistent multi-observable constraints on baryonic physics, including FRBs.
\vspace{.3cm}

\section*{Acknowledgements}
The authors would like to thank an anonymous referee for a very detailed report on this manuscript.
We would like to thank Ralf Konietzka for his help with the results from the TNG simulations. 
Furthermore, we would like to thank Zheng Zhang for many helpful conversations about emulators and MomentEMU. AN acknowledges support from the European Research Council (ERC) under the European Union’s Horizon 2020 research and innovation program with Grant agreement No. 101163128, as well as from the University of Manchester. SKG acknowledges the support by NWO (grant no OCENW.M.22.307) and Olle Engkvist Stiftelse (grant no. 232-0238). SH was supported by the Excellence Cluster ORIGINS, which is funded by the Deutsche Forschungsgemeinschaft (DFG, German Research Foundation) under Germany’s Excellence Strategy - EXC-2094 - 390783311. We used \texttt{matplotlib} \citep{Hunter_matplotlib_2007} for plots. A lot of the computations were done with the help of \texttt{SciPy} \citep{2020SciPy-NMeth} and \texttt{NumPy} \citep{harris2020array}. The computations in this work were performed using the Marvin HPC Cluster of the University of Bonn.

\clearpage

\vspace{.9cm}
\bibliographystyle{mnras}
\bibliography{MyLibrary,more_bib}

\renewcommand\thesection{\alpha{section}}
\renewcommand\thesubsection{\thesection.\arabic{subsection}}
\renewcommand\thefigure{\thesection.\arabic{figure}} 
\setcounter{figure}{0}

\appendix
\section{Choice of the maximum halo radius}
\label{app:rmax}

\begin{figure*}[t]
    \centering
    \includegraphics[width=\linewidth]{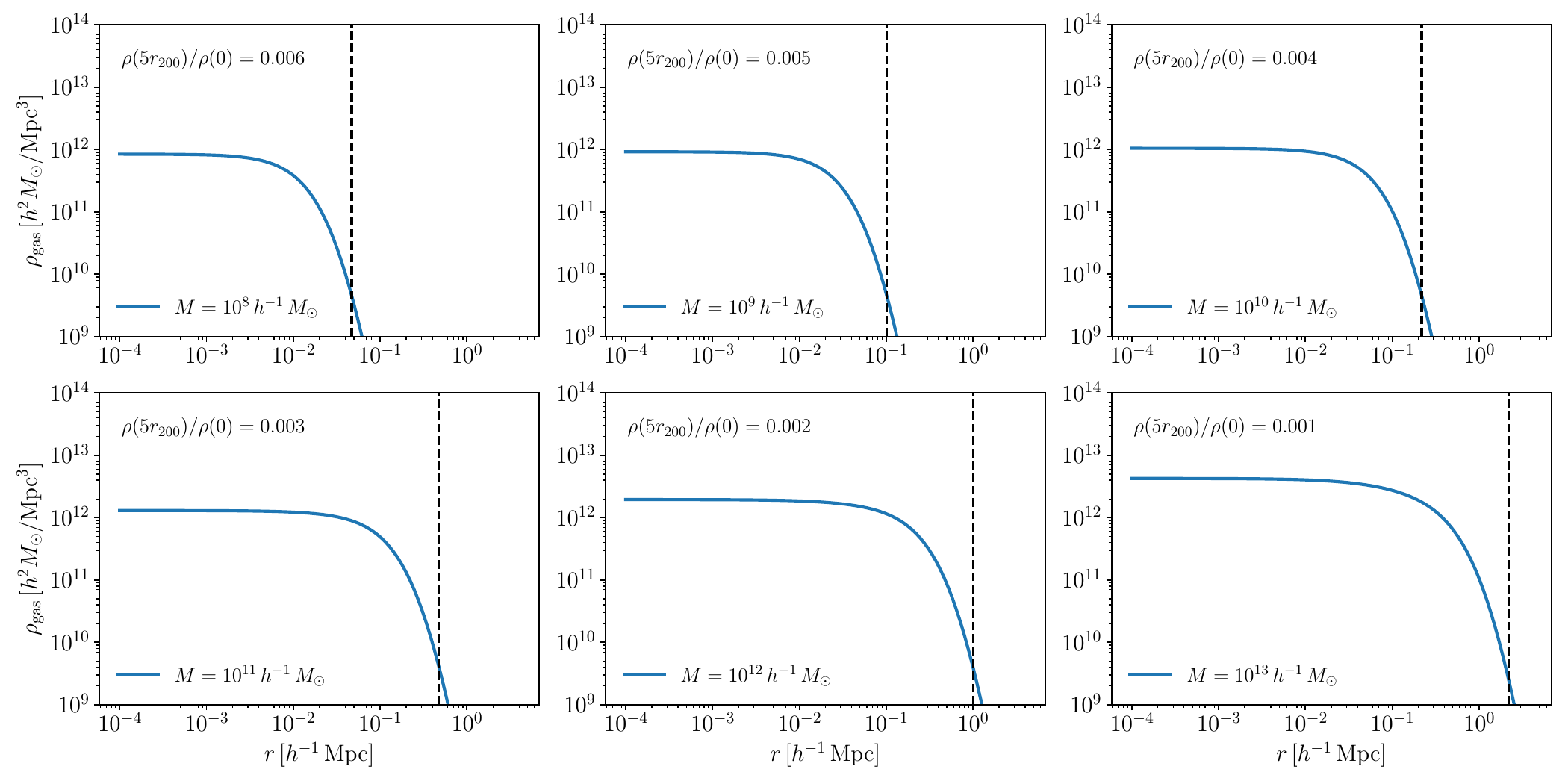}
    \caption{Gas density profiles for halos of different masses. The solid lines show the gas density profiles, while the dashed vertical lines mark $5r_{200}$ for each halo. The ratio between the gas density at $5r_{200}$ and the central value is indicated in the upper-left corner of each panel.}
    \label{fig:5rmax}
\end{figure*}

The maximum radius of an individual halo is generally not well defined and depends on the specific definition of a halo. The results of this paper depend strongly on this choice, since it determines the upper limit of the integration in \Cref{eq:DM_Halo}. To determine an appropriate maximum radius to use throughout this work, we begin by examining gas density profiles in different halo mass bins and identifying the radius at which the gas density around the centre drops significantly.

For this purpose, we consider a relatively flat gas density profile (e.g., a profile with a large $M_\mathrm{c}$ and a small $\delta$). We then choose a maximum radius at which the gas density has decreased by roughly a factor of $10^{3}$. Our calculations show that $5r_{200}$ is a safe choice for the maximum radius. The result is shown in \Cref{fig:5rmax}.

\begin{figure}
    \centering
    \includegraphics[width=\linewidth]{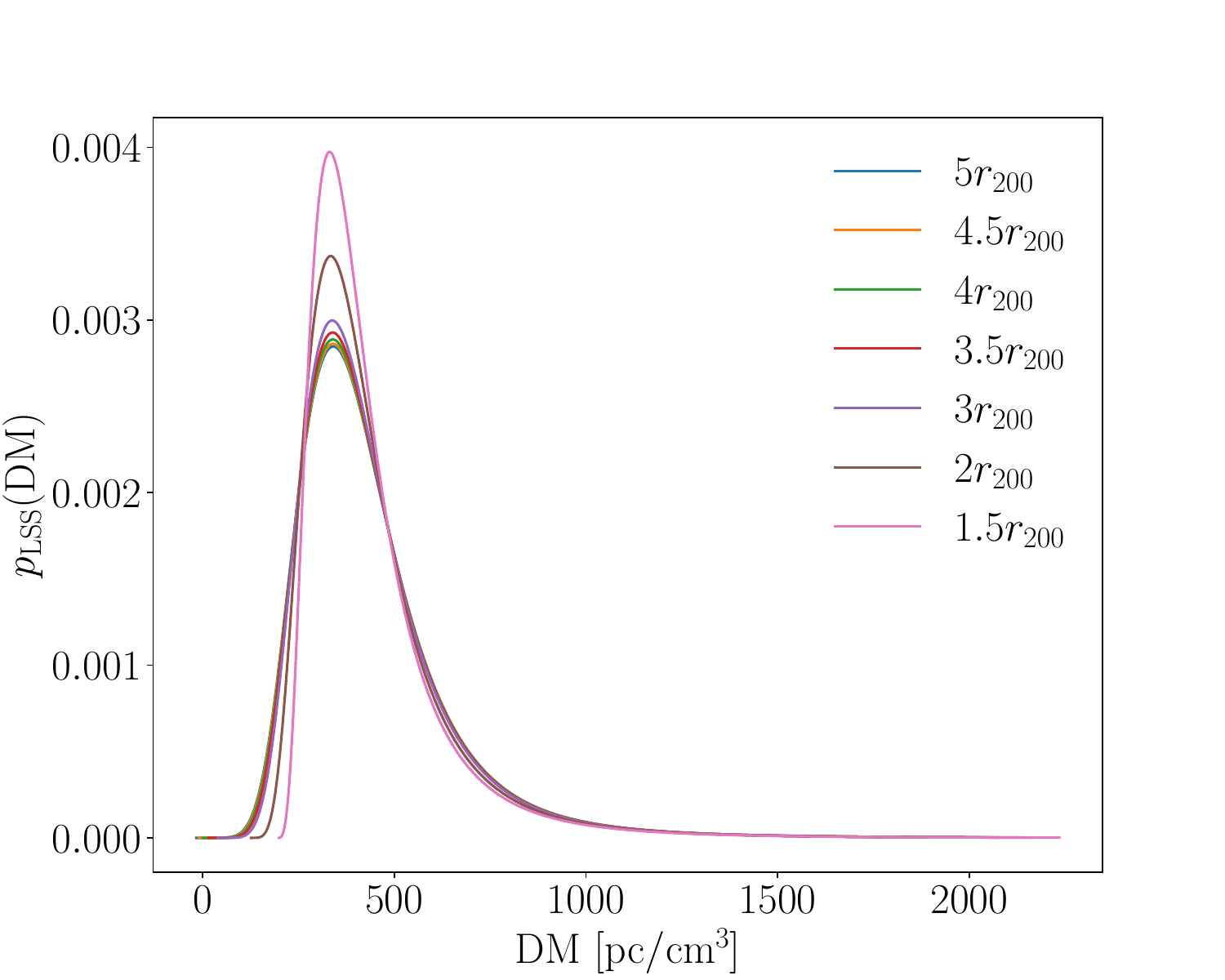}
    \caption{Comparison of the final PDFs obtained using different choices for the maximum halo radius.}
    \label{fig:r200}
\end{figure}

In addition, we investigate how this choice affects the final PDF. \Cref{fig:r200} shows how the PDF depends on the adopted maximum radius. As the maximum radius increases, the PDF becomes broader: it changes rapidly at small radii and then converges to a value near $4r_{200}$. This behaviour is expected. Increasing the maximum radius allows for larger DM values, while still producing small DM contributions from LOS passing near the outer edges of the halo. Beyond a certain radius, however, increasing the maximum radius no longer significantly changes the PDF, because the electron density decreases rapidly at large radii and these regions contribute negligibly to the total DM.

\section{\texttt{Cosmopower}}
\label{app:cosmopower}
To generate the training data set, we calculated 110,000 PDFs that cover a wide range of parameter
values. The parameter ranges used to generate this data are the same as the prior ranges listed in \Cref{tab:priorMCMC}, apart from $\delta\in [1,11]$
After training, we validate the emulator on a separate test dataset. Although the test parameters are within the same ranges as the training data set (as given in \Cref{tab:priorMCMC}), they are sampled independently and are not part of the training data set. For the \texttt{cosmopower}, we use \texttt{cosmopower\_PCAplusNN}, i.e. a principal component analysis compression before training the neural network. Four hidden layers, each with 512 nodes, are used, and the neural network is trained in 6 learning steps with a batch size of 1024.

\begin{figure}[h!]
    \centering    \includegraphics[width=.9\linewidth]{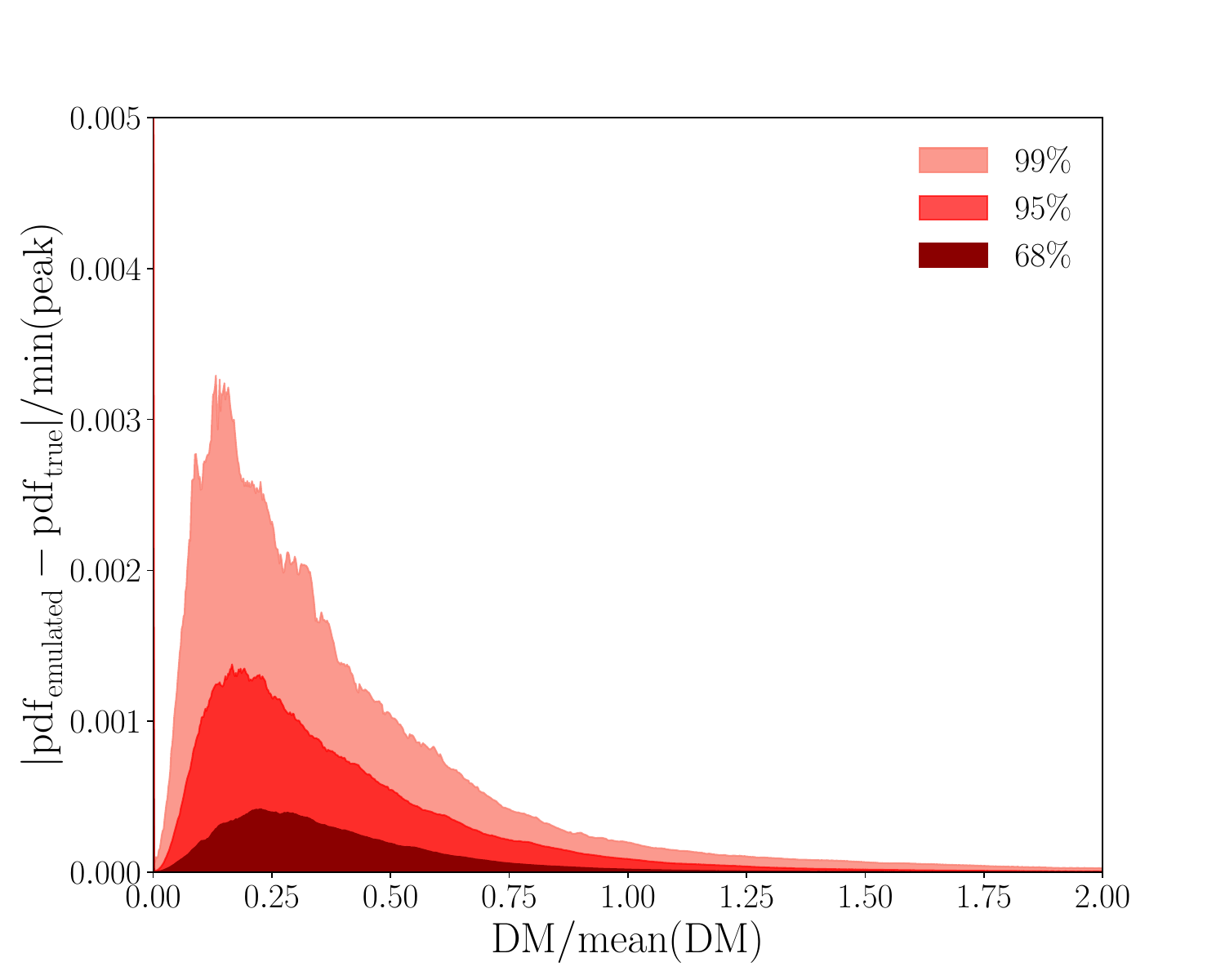}
    \caption{Accuracy of the emulator evaluated over the test data set. The plot shows the distribution of the relative difference between the emulated and true DM, normalised by the minimum of the two peaks. The shaded regions indicate that 68\%, 95\%, and 99\% of the test samples have relative differences below 0.001, 0.002, and 0.004, respectively, demonstrating high emulator precision.}
    \label{fig:emulator_test}
\end{figure}

Figure \ref{fig:emulator_test} illustrates the accuracy of the emulator. The x-axis represents the value of DM normalised by its mean at a given redshift, while the y-axis shows the relative difference between the emulated and true DM values, normalised by the minimum of their respective peaks. The plot demonstrates that 99\% of the test dataset exhibits a relative difference of less than 0.004. This result validates the emulator's performance.

\end{document}